\documentclass[aps,
superscriptaddress,amsmath,amssymb,showpacs,twocolumn]{revtex4-1}

\usepackage{graphicx}
\usepackage{hyperref}
\usepackage{epsfig}

\newcommand{\rem}[1]{}

\newcommand{\refe}[1]{~(\ref{#1})}

\newcommand{\Secref}[1]{Sec.~\ref{#1}}

\begin{document}

\title{Inelastic shot noise characteristics of nanoscale junctions from first principles}

\author{R. Avriller}
\affiliation{Univ.~Bordeaux, LOMA, UMR 5798, F-33400 Talence, France \\
CNRS, LOMA, UMR 5798, F-33400 Talence, France}
\affiliation{Donostia International Physics Center (DIPC) -- UPV/EHU, Paseo Manuel de Lardizabal 4,
E-20018 Donostia-San Sebastian, Spain}

\author{T. Frederiksen}
\affiliation{Donostia International Physics Center (DIPC) -- UPV/EHU, Paseo Manuel de Lardizabal 4,
E-20018 Donostia-San Sebastian, Spain}
\affiliation{IKERBASQUE, Basque Foundation for Science, E-48011, Bilbao, Spain}

\date{\today}

\begin{abstract}
We describe an implementation of \textit{ab-initio} methodology to compute inelastic shot noise signals 
due to electron-vibration scattering in nanoscale junctions.
The method is based on the framework of non-equilibrium Keldysh Green's functions
with a description of electronic structure and nuclear vibrations from density 
functional theory.
Our implementation is illustrated with simulations of electron transport in Au and Pt atomic point contacts.
We show that the computed shot noise characteristics of the Au contacts can be understood
in terms of a simple two-site tight-binding model representing the two apex atoms of the vibrating nano-junction. 
We also show that the shot noise characteristics of Pt contacts exhibit
more complex features associated with inelastic interchannel scattering.
These inelastic noise features are shown to provide additional information
about the electron-phonon coupling and the multichannel structure of Pt contacts 
than what is readily derived from the corresponding conductance characteristics.
We finally analyze a set of Au atomic chains of different lengths and strain conditions
and provide a quantitative comparison
with the recent shot noise experiments reported by Kumar \textit{et al.}~[Phys.~Rev.~Lett.~{\bf 108}, 146602 (2012)]. 
\end{abstract}


\maketitle

\section{Introduction}
\label{sec1}

The signatures of vibrational modes in the shot noise properties of nanoscale 
junctions have been the subject of active theoretical
investigations \cite{PhysRevB.53.10078,PhysRevB.67.165326,PhysRevB.71.045331,GaNiRa.06.Inelastictunnelingeffects,LuWa.07.Coupledelectronand,Schmidt:2009,Avriller:2009,Haupt:2009,Di-Ventra:2005,Urban:2010,Novotny:2011,Galperin:2011}.  
Recently, it was shown in Refs.~\cite{Schmidt:2009,Avriller:2009,Haupt:2009}, that under
applying a bias voltage $eV$ larger than the typical phonon energy $\hbar \omega_{0}$, the activation of 
phonon emission in a junction at low temperatures is responsible for a threshold behavior of the shot noise
versus voltage characteristics. 
More specifically, depending on the electronic transmission probability of the junction, the correction to the shot noise
signals induced by electron-phonon (e-ph) interactions was shown to exhibit jumps in
the voltage derivative that are either positive
or negative, as the result of a subtle interplay between one-electron tunneling events and correlated two-electrons
processes \cite{Kumar:2011}.
This behavior of the inelastic shot noise signal was recently shown to be strongly dependent on 
fluctuations in the occupation of the locally excited vibrational mode. Under certain conditions, 
this phenomenon might lead to a strong feedback of the dynamics of the oscillator on the electronic
noise properties \cite{Urban:2010,Novotny:2011} and the corresponding nonlinear effect in the shot noise
could be of interest for the characterization of heating effects at the nanoscale. 

In parallel to this theoretical activity, the first non-equilibrium shot noise measurements were
recently performed on gold (Au) nano-junctions that unravel clear signatures of the excitation of local vibrational modes 
\cite{Kumar:2011}.  
Those measurements confirm qualitatively the predictions of Ref.~\cite{Schmidt:2009,Avriller:2009,Haupt:2009} concerning
the existence of a crossover from positive to negative correction to the Fano factor upon phonon excitation,
when decreasing the electron transmission factor $\tau$ of the junction from unity. However, the 
experimental shot noise characteristics exhibit some unexplained features which seem to be out of the range of
``single level, single vibrational mode'' models.
For instance, the position of the crossover was shown to be shifted compared to the theory from 
$\tau \approx 0.85$ to $\tau \approx 0.95$ 
\cite{Kumar:2011}.

The above mentioned experiments exemplify the relevance of exploring new methods that 
allow to compute quantitatively the inelastic shot noise signals from first principles.
The aim of the present paper is thus two-fold: First, to document our implementation of such a framework into 
the \textsc{Inelastica} \cite{PaFrBr.05.Modelinginelasticphonon,Frederiksen:2007,Inelastica} \textit{ab-initio} code based 
on \textsc{Siesta} \cite{SIESTA:2002} and \textsc{TranSiesta} \cite{BrMoOr.02.Density-functionalmethodnonequilibrium}.
To this end, we adopt the results derived in Refs.~\cite{Komnik:2006,Haupt:2010} based on non-equilibrium 
Keldysh Green's functions.
Secondly, we apply our implementation to discuss the inelastic shot noise signals
in atomic point contacts of Au and Pt \cite{AgYeva.03.Quantumpropertiesof,UnCaCa.07.Formationofmetallic}
with all parameters extracted from atomistic calculations. 
Depending on the material, a different number of conductance channels are available for the
electron transport \cite{BrSoJa.97.Conductanceeigenchannelsin,ScAgCu.98.signatureofchemical,CuYeMa.98.MicroscopicOriginof}
and---as a consequence---also the inelastic transport properties are qualitatively
different \cite{BoEdSc.09.Point-contactspectroscopyaluminium}.
In particular, this allows us to highlight and analyze the additional complexities that 
arise from interchannel scattering under realistic conditions.
We finally analyze calculations for a set of Au atomic chains of different lengths and strain conditions
and compare the findings with the recent shot noise experiments \cite{Kumar:2011}.

The organization of the paper is the following. In \Secref{sec2}, we present the \textit{ab-initio} 
methodology we have implemented in order to compute the correction to the shot noise
induced by e-ph interactions.
In \Secref{sec3} we present results obtained for different Au and Pt atomic point contact,
and calculations for Au atomic chains bridging the electrodes is investigated in \Secref{sec4}. 
Our conclusions are presented in \Secref{sec5}.  

\section{Methodology}
\label{sec2}

In this section we outline the methodology used to perform first-principles calculations of 
shot noise characteristics.

\subsection{Model}
\label{sec2-1}
We consider the standard partitioning scheme in which an interacting 
device region $D$ couples to two reservoirs of noninteracting electrons, namely left $L$ and
right $R$ leads. This system is described by the following spin degenerate Hamiltonian
\begin{eqnarray}
\hat{H} &=& \hat{H}_D + \hat{H}_{L,R} + \hat{H}_T.
\end{eqnarray}
Here the device region $D$, in which the e-ph interactions are assumed to be strictly localized,
is described by a Hamiltonian of the form
\begin{eqnarray}
\hat{H}_D &=& \hat{H}^{(0)}_\textrm{e}+ \hat{H}^{(0)}_\textrm{ph}+\hat{H}_\textrm{e-ph},\\
\hat{H}^{(0)}_\textrm{e} &=& \sum_{i,j}H^{(0)}_{ij}{\hat d}^\dagger_{i}{\hat d}^{\phantom\dagger}_{j},
\label{eq:MFHamiltonian}\\
\hat{H}^{(0)}_\textrm{ph} &=& \sum_{\lambda}\hbar \omega_\lambda
{\hat b}^\dagger_{\lambda}{\hat
b}^{\phantom\dagger}_{\lambda},
\label{eq:HamFreeOscillators}\\
\hat{H}_\textrm{e-ph} &=& \sum_{\lambda}\sum_{i,j}
M^{\lambda}_{ij}{\hat d}^\dagger_{i}{\hat
d}^{\phantom\dagger}_{j} ({\hat b}^\dagger_{\lambda}+{\hat
b}^{\phantom\dagger}_{\lambda}),\label{eq:ephCouplingHamiltonian}
\end{eqnarray}
where ${\hat d}^\dagger_{i}$ and ${\hat
b}^\dagger_{\lambda}$ are the electron and phonon creation
operators in the device space, respectively.
Here $\hat{H}^{(0)}_\textrm{e}$ is the
single-particle Kohn Sham DFT Hamiltonian describing electrons moving
in a static arrangement of the atomic nuclei. In this Hamiltonian, electron-electron interactions 
are taken into account at the mean field level. 
$\hat{H}^{(0)}_\textrm{ph}$ is the phonon Hamiltonian of free uncoupled
harmonic oscillators, and $\hat{H}_\textrm{e-ph}$ is the e-ph
coupling within the harmonic approximation.

The Hamiltonians describing the leads $\hat{H}_{L,R}$ and the tunnel couplings between leads and 
device region $\hat{H}_T$ are given by 
\begin{eqnarray}
\hat{H}_{L,R} &=& \sum_{\alpha=L,R}\sum_{i,j}H^\alpha_{ij}{\hat c}^\dagger_{\alpha,i}{\hat c}^{\phantom\dagger}_{\alpha,j},\\
\hat{H}_{T} &=& \sum_{\alpha=L,R}\sum_{i,j}\left(V^\alpha_{ij}{\hat c}^\dagger_{\alpha,i}{\hat d}^{\phantom\dagger}_{j}+h.c.\right),
\end{eqnarray}
where ${\hat c}^\dagger_{\alpha,i}$ is the electron creation operator in lead $\alpha=L,R$.
Each lead is considered to be in local equilibrium such that the occupied states
are characterized by a Fermi distribution with thermal energy $k_BT$ and chemical potential 
$\mu_\alpha$. An applied voltage $V$ is assumed to shift symmetrically the chemical potentials
of the leads with respect to the Fermi level position at equilibrium $E_{F}$
(determined self-consistently in Kohn-Sham DFT by filling the electronic states from below)
as $\mu_{L(R)}= E_{F} +(-) eV/2$ and to leave the electrostatic 
potential of the device unchanged.

\subsection{Electronic structure methods}
\label{sec2-2}
All parameters in the above Hamiltonian are extracted from self-consistent 
calculations with \textsc{Siesta} \cite{SIESTA:2002} for the \textsc{TranSiesta} setup \cite{BrMoOr.02.Density-functionalmethodnonequilibrium}
according to the \textsc{Inelastica} scheme \cite{PaFrBr.05.Modelinginelasticphonon,Frederiksen:2007,Inelastica}.
With $\{\phi_i\}$ denoting the full nonorthogonal basis set of atomic orbitals, the
Fermion operators satisfy the anti-commutation relations 
$\{{\hat d}^\dagger_{i},{\hat d}^\dagger_{j}\}=\{{\hat d}_{i},{\hat d}_{j}\}=0$ and
$\{{\hat d}_{i},{\hat d}^\dagger_{j}\}=S_{ij}$, where $S_{ij}=\langle \phi_i|\phi_j\rangle$ is the overlap matrix
(and similar for the lead Fermion operators). 
Using boldface notation throughout this manuscript for the electronic space, we let
$\mathbf H^{(0)} $ represent the matrix elements of the Kohn-Sham Hamiltonian in the device region,
$\mathbf V^\alpha$ the coupling elements between device and lead $\alpha$, $\mathbf S$ the overlap
matrix, and $\mathbf M^\lambda$ the e-ph coupling matrix (obtained by finite differences \cite{Frederiksen:2007}) corresponding 
to a localized mode $\lambda$ with energy $\hbar\omega_\lambda$.

In the device region $D$ the single-particle noninteracting retarded (advanced) Green's function $\mathbf{g}^{r(a)}(E)$, 
\textit{i.e.}, without e-ph interactions, takes the usual form
\begin{eqnarray}
\mathbf{g}^{r(a)}(E) &=& \{(E\pm i0^+) \mathbf S-\mathbf H^{(0)}-\mathbf\Sigma_{L}^{r(a)} (E)-\mathbf\Sigma_{R}^{r(a)} (E)\}^{-1},\nonumber\\
\end{eqnarray}
where 
\begin{eqnarray}
\mathbf\Sigma_\alpha^{r(a)}(E) &=& (\mathbf  V^\alpha)^\dagger \mathbf g^{r(a)}_{\alpha S} (E) \mathbf V^\alpha
\end{eqnarray}
is the retarded (advanced) self-energy due to lead $\alpha$. Here
$\mathbf g^{r(a)}_{\alpha S}(E)$ represents the corresponding surface Green's function 
for the isolated lead and is calculated recursively \cite{SaSaRu.85.Highlyconvergentschemes}. 
The retarded and advanced Green's functions are connected through the 
relation $\mathbf{g}^{a}(E) = \{\mathbf{g}^{r}(E)\}^\dagger$.
For convenience we also introduce here the level broadening due 
to lead $\alpha$
\begin{eqnarray}
\mathbf\Gamma_{\alpha}(E) &=& i\lbrace\mathbf\Sigma_{\alpha}^{r}(E) - \mathbf\Sigma_{\alpha}^{a}(E)\rbrace.
\end{eqnarray}

Finally, as the theory of shot noise characteristics presented in the following section
is developed in an orthogonal basis, we orthogonalize all relevant quantities according
to standard L\"owdin transformations \cite{Lo.50.NonOrthogonalityProblem}, \textit{i.e.},
\begin{eqnarray}
\mathbf H^{(0)} &\rightarrow& \mathbf {\widetilde H}^{(0)} = \mathbf S^{-1/2} \mathbf H^{(0)} \mathbf S^{-1/2}, \label{eq:ortho-first}\\
\mathbf M^\lambda &\rightarrow& \mathbf {\widetilde M}^\lambda = \mathbf S^{-1/2} \mathbf M^\lambda\mathbf S^{-1/2}, \\
\mathbf S &\rightarrow& \mathbf {\widetilde S} = \mathbf S^{-1/2} \mathbf S \mathbf S^{-1/2} = \mathbf 1,\\
\mathbf g^{r(a)} &\rightarrow& \mathbf {\widetilde g}^{r(a)} = \mathbf S^{1/2} \mathbf g^{r(a)} \mathbf S^{1/2}, \\
\mathbf \Gamma_\alpha &\rightarrow& \mathbf {\widetilde\Gamma}_\alpha = \mathbf S^{-1/2} \mathbf \Gamma_\alpha \mathbf S^{-1/2},\label{eq:ortho-last}
\end{eqnarray}
where the transformation matrices
\begin{eqnarray}
\mathbf S^{1/2} &=& \mathbf U\mathrm{diag}(\sqrt{\varepsilon_1},\ldots,\sqrt{\varepsilon_n}) \mathbf U^{-1},\\
\mathbf S^{-1/2} &=& \mathbf U \mathrm{diag}(1/\sqrt{\varepsilon_1},\ldots, 1/\sqrt{\varepsilon_n}) \mathbf U^{-1},
\end{eqnarray}
are determined through the eigenvalue problem $\mathbf S U_i = \varepsilon_i U_i$ for
the overlap matrix.

\subsection{Nonequilibrium Keldysh Green's functions}
\label{sec2-3}
We summarize in this section how to calculate conductance and shot noise characteristics for
nano-junctions described by the electronic structure as outlined in \Secref{sec2-2}. 
The concepts underlying this methodology are derived in Ref.~\cite{Komnik:2006} and were applied 
to single-level and single-mode models in Refs.~\cite{Schmidt:2009,Avriller:2009,Haupt:2009}.
Here we follow the specific generalization to the multi-level and multi-mode model formulated 
by Haupt {\it et.~al.} \cite{Haupt:2009,Haupt:2010}. The two fundamental approximations are
(i) weak e-ph interactions and (ii) the so-called extended wide-band limit (EWBL).

\subsubsection{Extended wide-band limit (EWBL)}
As in previous work we adopt the EWBL 
\cite{PaFrBr.05.Modelinginelasticphonon,ViCuPa.05.Electron-vibrationinteractionin,
VeMaAg.06.Universalfeaturesof,Frederiksen:2007,Haupt:2009,Haupt:2010} to perform
energy integrations analytically such that explicit results for the mean current and shot noise
can be stated. The EWBL consists of approximating the noninteracting retarded and advanced Green's functions 
as well as the level broadenings with their values at the Fermi energy $E_F$, \textit{i.e.},
\begin{eqnarray}
\mathbf{g}^{r(a)}(E) &\approx& \mathbf{g}^{r(a)}(E_F) \equiv \mathbf{g}^{r(a)} ,\label{eq:eWBL1}\\
\mathbf\Gamma_{\alpha}(E) &\approx& \mathbf\Gamma_{\alpha}(E_F) \equiv \mathbf\Gamma_{\alpha} .\label{eq:eWBL2}
\end{eqnarray}
Physically this is motivated by the fact that in many real systems the electronic spectral 
properties typically vary slowly on the scale of a few phonon energies and applied voltages
\cite{PaFrBr.05.Modelinginelasticphonon,ViCuPa.05.Electron-vibrationinteractionin,
VeMaAg.06.Universalfeaturesof,Frederiksen:2007}. In the case of atomic gold wires this approximation was
successfully tested by one of us in Ref.\cite{Frederiksen:2007} via a direct comparison to computationally
more expensive calculations based on the self-consistent Born approximation. The physical reason is the 
strong hybridization of device states with the electrode states (life-time broadening on the eV scale) and 
the existence of only low-energy vibrational modes (on the meV scale). As this situation also applies to
point contacts of Au and Pt, we expect the EWBL to be a very good approximation for these systems too.

\subsubsection{Computing the current characteristics}
\label{sec2-2-b}

In absence of e-ph interactions ($\mathbf{M}^\lambda=\mathbf{0}$), the (bare) current $I_{0}(V)$ is given by the standard 
Landauer-B\"uttiker formula \cite{Buttiker:1986}. Within EWBL the
transmission function becomes energy independent
so that the expression for the current-voltage characteristics is simply   
\begin{eqnarray}
I_{0}[2e/h](V) &=& \mbox{Tr}\lbrace \mathbf{T} \rbrace eV,   
\label{eqn:Current_No_Interaction}
\end{eqnarray}  
where
\begin{eqnarray}
\mathbf{T} &=& \mathbf{\Gamma}_{L}\mathbf{g}^{r}\mathbf{\Gamma}_{R}\mathbf{g}^{a}
\label{eqn:Transmission_Factor} 
\end{eqnarray}
is the elastic (bare) transmission matrix of the 
junction. This expression is similar to the Fisher and Lee formula for
the conductance \cite{Fisher:1981}.
In presence of weak coupling to the vibrational subsystem ($\mathbf{M}^\lambda \ne \mathbf{0}$),
the above expression for the mean electronic current has to be modified. At second order of perturbation 
theory in the e-ph coupling strength, the correction to the current $\delta I(V)$ (within EWBL) can be expressed 
in terms of products of microscopic factors (system dependent) by voltage-dependent universal 
functions (system independent) \cite{Frederiksen:2007,Haupt:2010}. 
We write the inelastic corrections to the current as
\begin{eqnarray}
\delta I[2e/h](V) &=& \delta I_\mathrm{el}(V) + \delta I_\mathrm{inel}(V) , 
\label{eqn:Current_Interaction} \\
\delta I_\mathrm{el}[2e/h](V) &=& \sum_\lambda\Big{\lbrace} (1+2n_{B}^\lambda) \mbox{Tr}\lbrace \mathbf{T}^\mathrm{(0)}_\lambda\rbrace
\label{eqn:Current_Interaction_Elastic_Part} \\
 &&+ 2n_{B}^\lambda \mbox{Tr}\lbrace\mathbf{T}^\mathrm{(1)}_\lambda\rbrace \Big{\rbrace} eV   ,
\nonumber \\
\delta I_\mathrm{inel}[2e/h](V) &=& \sum_\lambda\mbox{Tr}\lbrace \mathbf{T}^\mathrm{(1)}_\lambda \rbrace g_\lambda(eV)   .
\label{eqn:Current_Interaction_Inelastic_Part}  
\end{eqnarray}  
with the microscopic factors given by traces over the following quantities
\begin{eqnarray}
\mathbf{T}^\mathrm{(0)}_\lambda &=& \mathbf{\Gamma}_{L} \big{(} \mathbf{g}^{r}\mathbf{M}^\lambda\mathbf{g}^{r}_\mathrm{Re}
\mathbf{M}^\lambda\mathbf{A}_{R} + \mbox{H.c.} \big{)},
\label{eqn:Elastic_Transmission_Factor} \\
\mathbf{T}^\mathrm{(1)}_\lambda &=& \mathbf{\Gamma}_{L} \mathbf{g}^{r} \big{\lbrace} \mathbf{M}^\lambda\mathbf{A}_{R}\mathbf{M}^\lambda
\label{eqn:Inelastic_Transmission_Factor} \\
&&- \frac{i}{2}  \big{(}  \mathbf{M}^\lambda\mathbf{A}\mathbf{M}^\lambda\mathbf{g}^{r}\mathbf{\Gamma}_{R} -
\mbox{H.c.} \big{)} \big{\rbrace} \mathbf{g}^{a} , \nonumber
\end{eqnarray}
and the voltage dependent universal functions given by
\begin{eqnarray}
 g_\lambda(eV) &=& eV + \frac{1}{2} \big{\lbrace} U(eV-\hbar\omega_\lambda) - U(eV+\hbar\omega_\lambda)\big{\rbrace},\quad
\label{eqn:Universal_function_g} \\
 U(eV) &=& eV \coth{(\beta eV/2)}.
\label{eqn:Universal_function_U}
\end{eqnarray}  
In the above equations, $\mathbf{g}^{r}_\mathrm{Re}$ denotes the real part of $\mathbf{g}^{r}$, $\beta=1/k_BT$ the inverse temperature of the electrodes, and
$\mathbf{A}_\alpha=\mathbf{g}^{r}\mathbf{\Gamma}_\alpha\mathbf{g}^{a}$ the partial spectral function corresponding to
lead $\alpha$. The total spectral function is given by $\mathbf{A} = \mathbf{A}_{L} + \mathbf{A}_{R}$.

We note that the expressions Eqs.~(\ref{eqn:Current_Interaction_Elastic_Part})-(\ref{eqn:Current_Interaction_Inelastic_Part}) 
are subject to the following three additional approximations: 
(i) We have ignored the asymmetric contributions to the conductance (with respect to voltage)
which are derived in Ref.~\cite{Egger:2008,Entin-Wohlman:2009,Haupt:2010}. These contributions are 
logarithmically divergent in the zero-temperature limit for $|eV|=\hbar\omega_\lambda$ and signal the breakdown of 
second-order perturbation theory at the inelastic threshold. However, a resummation scheme might 
renormalize and cure this problem \cite{Urban:2010}.
Out of the threshold region, the logarithmic corrections to the conductance are typically
orders of magnitude smaller than the symmetric contribution. They also vanish in the 
limit of symmetric couplings to the left and right lead \cite{Frederiksen:2007}, thus justifying our 
assumption.
(ii) We fix the phonon populations $n_B^\lambda$ to the equilibrium values as given by the Bose-Einstein distribution
$n_B^\lambda=1/[\exp(\beta \hbar\omega_\lambda)-1]$ 
(regime of equilibrated phonons). For atomic point contacts and chains this is a reasonable starting
point as the vibrations in the nanoscale contact are damped to some extent by coupling to bulk phonons
\cite{FrBrLo.04.InelasticScatteringand,PaFrBr.05.Modelinginelasticphonon,Frederiksen:2007,EnBrJa.09.Atomistictheorydamping}.
Furthermore, the effect of phonon heating on the shot noise characteristics is a delicate research topic
that is beyond the scope of the present study \cite{Urban:2010,Novotny:2011,Galperin:2011}.
(iii) Equation \refe{eqn:Current_Interaction} corresponds to taking into account only the Fock (exchange) self-energy. 
Neglecting the Hartree term is indeed a good approximation in most cases of interest: within the EWBL
this term provides a voltage independent renormalization of the molecular level positions in the regime of equilibrated phonons 
and thus displays no features at the phonon emission threshold \cite{Haupt:2010}.

The total correction to the current in Eq.\refe{eqn:Current_Interaction} is 
the sum of two contributions.
The first one $\delta I_\mathrm{el}(V)$ is due to elastic processes induced by e-ph interactions that
renormalize the bare transmission factor $\tau$ to an effective one, e.g., 
$\tilde{\tau} \approx \tau + \sum_\lambda\mbox{Tr}\lbrace \mathbf{T}^\mathrm{(0)}_\lambda\rbrace$ at zero temperature. 
The second contribution $\delta I_\mathrm{inel}(V)$ originates from inelastic
processes activated by phonon emission. The voltage dependence of this contribution is a universal 
function of voltage $g_\lambda(V)$ [see Eq.\refe{eqn:Universal_function_g}] that exhibits a threshold at $|eV|=\hbar\omega_\lambda$,
in the zero temperature limit. The sign and size of this threshold (jump in conductance) is controlled by
the inelastic microscopic factors $\mathbf{T}^\mathrm{(1)}_\lambda$ given in Eq.\refe{eqn:Inelastic_Transmission_Factor}.
Upon differentiation of Eq.\refe{eqn:Current_Interaction} with respect to voltage one obtains
the correction of the conductance $\delta G(V)=\partial_{V}(\delta I(V))$. In the zero temperature limit, 
$\delta G(V)$ is discontinuous at the inelastic threshold, due 
to the contribution of the inelastic term, see Eq.~\refe{eqn:Current_Interaction_Inelastic_Part}. 
We thus define the corresponding jump in the inelastic correction to 
the conductance $\Delta G_\lambda \equiv \lim_{\eta\rightarrow 0^{+}} \lbrace \delta G(\hbar\omega_\lambda+\eta) -
\delta G(\hbar\omega_\lambda-\eta)\rbrace$ at the threshold voltage corresponding to mode $\lambda$ by

\begin{eqnarray}
\Delta G_\lambda[2e^{2}/h] = \mbox{Tr}\lbrace \mathbf{T}^\mathrm{(1)}_\lambda \rbrace   .
\label{eqn:Jump_Conductance_Tot}
\end{eqnarray}

\subsubsection{Computing the shot noise characteristics}
\label{sec2-2-c}
Shot noise characteristics in absence of coupling to local vibrational modes ($\mathbf{M}^\lambda = \mathbf{0}$) was reviewed
by Blanter and B\"{u}ttiker in Ref.~\cite{Blanter:2000}. Within EWBL, the correlation
function of the current operator evaluated at zero frequency, e.g. the (bare) shot noise 
characteristics $S_{0}(V)$, is given by the simplified expression \cite{Blanter:2000} 
\begin{eqnarray}
S_{0}[2e^{2}/h](V) &=& \frac{2}{\beta} \mbox{Tr}\lbrace \mathbf{T}^{2} \rbrace + 
\mbox{Tr}\lbrace \mathbf{T}( \mathbf{1} - \mathbf{T} ) \rbrace U(eV).\quad   
\label{eqn:Noise_No_Interaction}
\end{eqnarray}  
This expression is associated with the noise induced by thermal fluctuations 
in the electron occupation of the
electrode Fermi seas and with fluctuations in the occupation 
of the coherent left- and right-moving scattering states.

At second order of perturbation theory in the e-ph coupling strength,
the expression for the finite temperature correction to the noise $\delta S(V)$
in the regime of equilibrated phonons
is rather complicated, and derived in detail in Ref.~\cite{Haupt:2010}.
While we have implemented these results
and used them in the numerical part of this
work, we here just state the simpler result for the zero-temperature limit

\begin{eqnarray}
\delta S[2e^{2}/h](V) &=& \delta S_\mathrm{el}(V) + \delta S_\mathrm{inel}(V)  ,
\label{eqn:Noise_Interaction} \\
\delta S_\mathrm{el}[2e^{2}/h](V) &=&  \sum_\lambda\mbox{Tr}\Big{\lbrace} ( \mathbf{1} - 2\mathbf{T} )\mathbf{T}^\mathrm{(0)}_\lambda \Big{\rbrace} |eV| ,
\label{eqn:Noise_Interaction_Elastic_Part} \\
\delta S_\mathrm{inel}[2e^{2}/h](V) &=& \sum_\lambda\mbox{Tr}\Big{\lbrace} ( \mathbf{1} - 2\mathbf{T} )\mathbf{T}^\mathrm{(1)}_\lambda
\label{eqn:Noise_Interaction_Inelastic_Part}
+ \mathbf{Q}_\lambda \Big{\rbrace} \qquad\\
&&\times (|eV| - \hbar\omega_\lambda)\theta(|eV| - \hbar\omega_\lambda)   ,
\nonumber  
\end{eqnarray}
with
\begin{eqnarray}
\mathbf{Q}_\lambda &=& - \mathbf{g}^{a}\mathbf{\Gamma}_{L}\mathbf{g}^{r} \big{\lbrace}
\mathbf{M}^\lambda\mathbf{A}_{R}\mathbf{\Gamma}_{L}\mathbf{A}_{R}\mathbf{M}^\lambda
\label{eqn:Inelastic_Q_Factor} \\
&&\quad+ \mathbf{M}^\lambda\mathbf{A}_{R}\mathbf{\Gamma}_{L}\mathbf{g}^{r}\mathbf{M}^\lambda\mathbf{g}^{r}\mathbf{\Gamma}_{R} +
\mbox{H.c.} \big{\rbrace} .
\nonumber 
\end{eqnarray}  
Consistent with our assumptions for the inelastic corrections to the mean current (\Secref{sec2-2-b})
we neglect also for the noise part the asymmetric terms leading to logarithmic divergences as well as 
contributions from the Hartree e-ph self-energy.

Analogous to the corrections to the current [Eqs.~(\ref{eqn:Current_Interaction})-(\ref{eqn:Universal_function_U})] also
the inelastic noise corrections [Eqs.~(\ref{eqn:Noise_Interaction})-(\ref{eqn:Inelastic_Q_Factor}) in the zero-temperature limit] 
can be written as products of microscopic factors by universal voltage-dependent functions. 
The first term $\delta S_\mathrm{el}(V)$ [Eq.\refe{eqn:Noise_Interaction_Elastic_Part}] represents an elastic correction to the noise.
The second term $\delta S_\mathrm{inel}(V)$ [Eq.\refe{eqn:Noise_Interaction_Inelastic_Part}] is related to inelastic
signatures of phonon activation in the shot noise. Its origin and interpretation is less intuitive
than the corresponding expression for the mean current.
The part proportional to  
$\mbox{Tr}\{( \mathbf{1} - 2\mathbf{T} )\mathbf{T}^\mathrm{(1)}_\lambda\}$ originates
from a mean-field contribution to the shot noise while
the other part proportional to
$\mbox{Tr}\{\mathbf{Q}_\lambda \}$ is related to vertex corrections
\cite{Haupt:2010}.
A similar (although slightly different) decomposition 
in terms of one-electron (mean-field like) and two electron (vertex-like) processes
was proposed in Ref.~\cite{Kumar:2011}.

Instead of looking directly at the inelastic corrections in the shot noise 
it is convenient to analyze the voltage derivative of the shot noise $\delta \dot{S}(V)=\partial_{V}(\delta S(V))$,
\textit{i.e.}, the inelastic noise change.
In the zero temperature limit, 
$\delta \dot{S}_\lambda(V)$ is discontinuous at the inelastic threshold, due 
to the contribution of the inelastic term, see Eq.\refe{eqn:Noise_Interaction_Inelastic_Part}.
We thus define the corresponding jump in the inelastic correction to 
the shot-noise $\Delta \dot{S}_\lambda \equiv \lim_{\eta\rightarrow 0^{+}} \lbrace \delta \dot S(\hbar\omega_\lambda + \eta) -
\delta \dot S(\hbar\omega_\lambda - \eta) \rbrace$ at the threshold voltage corresponding to mode $\lambda$ by

\begin{eqnarray}
\Delta \dot{S}_\lambda[2e^{3}/h]= \mbox{Tr}\Big{\lbrace} ( \mathbf{1} - 2\mathbf{T} )\mathbf{T}^\mathrm{(1)}_\lambda
+ \mathbf{Q}_\lambda \Big{\rbrace}  .
\label{eqn:Jump_Noise_Tot}
\end{eqnarray}  

\section{Results for Au and Pt contacts}
\label{sec3}

%
\begin{figure}[ht]
\centering
   \includegraphics[width=\columnwidth]{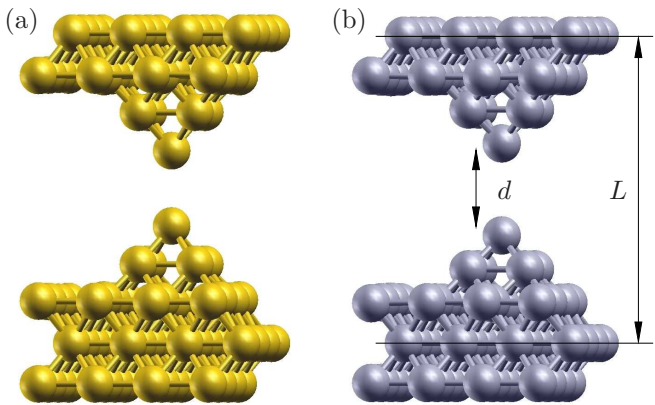}
\caption{\label{AuPt-contacts:fig} 
(Color online) Atomic point contacts of (a) Au atoms (shown in yellow-gold) and (b) Pt 
atoms (shown in blue-gray) considered in the first-principles transport calculations.
The characteristic electrode separation $L$ is measured between the second-topmost
surface layers.
The distance $d$ characterizes the separation between the two apex atoms.
}
\end{figure}
%
%
We first consider Au and Pt atomic point contacts, such as those shown Fig.~\ref{AuPt-contacts:fig},
as benchmark systems for our \textit{ab-initio} calculations.
Along the lines of Ref.~\cite{FrLoPa.07.Fromtunnelingto} by one of us,
we consider periodic supercells with a 4$\times$4 representation of 
either Au(100) and Pt(100) surfaces sandwiching two pyramids pointing toward each other.
The characteristic electrode separation $L$ is measured between the second-topmost surface 
layers, since the surface layers themselves are relaxed and hence deviate on the decimals from 
the bulk values.

Our \textsc{Siesta} calculations use a single-$\zeta$ plus polarization (SZP) 
basis with a confining energy of 0.01 Ry [corresponding to the 5$d$ and 6($s,p$) 
states of the free atoms], the generalized gradient approximation (GGA) for exchange-correlation, 
a cutoff energy of 200 Ry for the real-space grid integrations, and the $\Gamma$-point
approximation for the sampling of the three-dimensional Brillouin zone.
The interaction between the valence electrons and the ionic cores is described by standard
norm-conserving Troullier-Martins pseudopotentials generated from relativistic atomic calculations.
As for the bulk Au and Pt crystals, we set the lattice constants to 4.18 {\AA} 
and 4.02 {\AA}, respectively.

For each electrode separation $L$ we relax the surface atoms until the 
residual forces are smaller than 0.02 eV/{\AA} and proceed calculating 
vibrational modes and e-ph couplings by finite differences.
For simplicity, we here only consider that the two apex atoms can vibrate, leaving us
with six characteristic vibrational modes in the device. This assumption is only made to facilitate
a fundamental understanding of the inelastic signals.
Finally, while electron transport in the supercell approach generally involves a 
sampling over $k_{||}$-points, we approximate in the following all relevant quantities 
with their values at the $\Gamma$-point. 
%
\begin{figure}[ht]
\centering
   \includegraphics[width=\columnwidth]{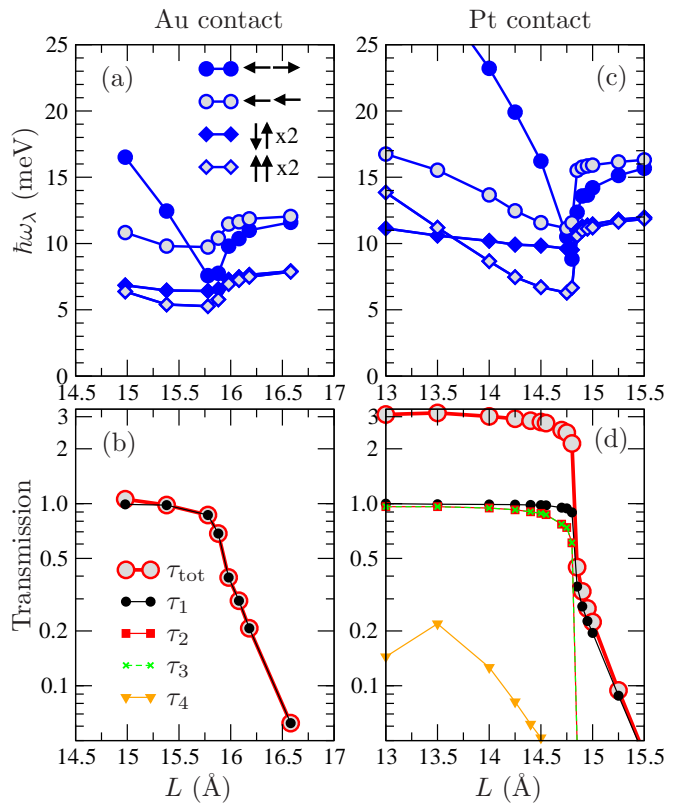}
\caption{\label{FreqTrans_vs_L:fig} 
(Color online) (a) Vibrational frequencies and (b) electronic transmission at the Fermi energy
as a function of electrode separation $L$ for the Au atomic point contact. (c),(d) similar
as (a),(b) but for the Pt atomic point contact.
As indicated in the legends in panel (a) the longitudinal (transverse) 
eigenmodes are pictured in (a) and (c) with circular (diamond) symbols. 
The out-of-phase (in-phase) modes are represented by filled (open) symbols.
In panels (b) and (d) both the total transmission $\tau=\tau_\mathrm{tot}$ (large red circles) 
as well as the transmission $\tau_{1},\ldots,\tau_{4}$ for the four most conducting transport eigenchannels (small symbols)
are shown.
Note that the legends shown in the Au panels apply also to the Pt panels.}

\end{figure}
%
%
\begin{figure*}[ht]
   \includegraphics[width=2\columnwidth]{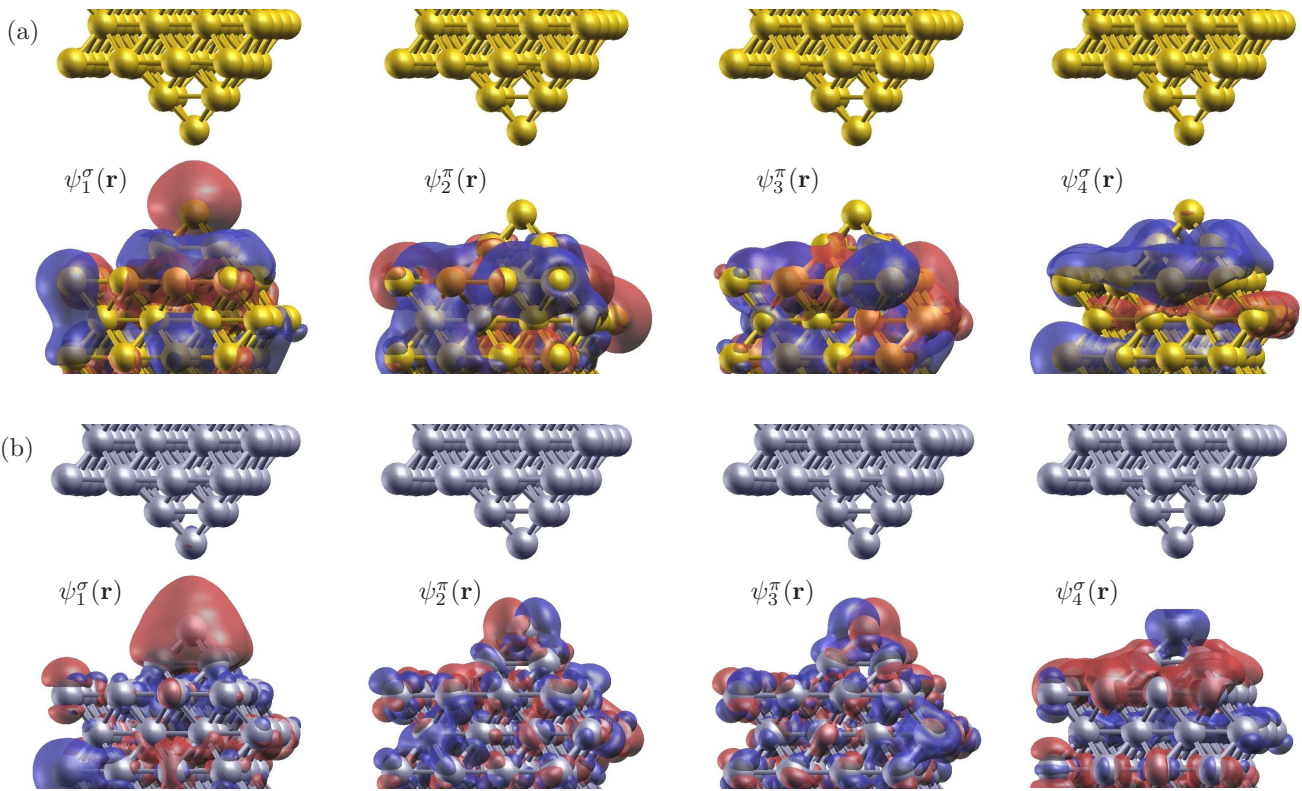}
\caption{\label{Eigenchannels:fig}
(Color online) Isosurface representations of the four most conducting eigenchannel scattering states $\psi_i(\mathbf{r})$ 
(incoming from below) for (a) a Au atomic point contact ($L= 16.58$ {\AA}) and (b) a Pt atomic point contact ($L=15.50$ {\AA}). The states are ordered according to decreasing transmission.
Due to the tunnel gap between the two sides the electron scattering states decay
rapidly and the transmitted part of the wave on the other electrode is not visible. The blue and red colors represent
the sign of the real part of the scattering states (our choice of phase makes the imaginary part negligible for visualization purposes).
The isosurface plots reveal different rotational symmetry around an axis connecting the apex atoms.
For both contacts one observes that the channels $\psi_1^\sigma$ and $\psi_4^\sigma$ are $\sigma$-type states (rotationally symmetric)
while $\psi_2^\pi$ and $\psi_3^\pi$ are $\pi$-type states (with a nodal plane through the symmetry axis).}
\end{figure*}
%
%

We present in Fig.~\ref{FreqTrans_vs_L:fig} the dependence of the electron transmission
and vibrational frequencies, as a function of the electrode separation $L$.
In both cases of Au and Pt junctions, the total transmission $\tau$ decreases with $L$
as observed in Fig.~\ref{FreqTrans_vs_L:fig}(b),(d) and covers the range from 
contact (ballistic limit) to the tunnel regime (low-transmission regime). 
The total transmission $\tau=\sum_i\tau_i$ can be understood as a sum over eigenchannel 
transmissions $\tau_i$ for a set of (non mixing) electron scattering states.
In Fig.~\ref{Eigenchannels:fig} we have visualized the scattering states belonging
to the four most transmitting channels (waves incoming from below) \cite{PaBr.07.Transmissioneigenchannelsfrom}.

In the case of Au junctions the total transmission is essentially made up of a single 
channel, \textit{i.e.}, $\tau\approx\tau_1$ as seen in Fig.~\ref{FreqTrans_vs_L:fig}(b).
This fact can be traced back to the single $s$-valence of Au
\cite{BrSoJa.97.Conductanceeigenchannelsin,ScAgCu.98.signatureofchemical,CuYeMa.98.MicroscopicOriginof,AgYeva.03.Quantumpropertiesof}.
The corresponding eigenchannel scattering state $\psi_{1}^{\sigma}$ is rotationally symmetric 
($\sigma$-type) as seen in Fig.~\ref{Eigenchannels:fig}(a). The fact that essentially
only one transmission channel contributes to the elastic current can also be appreciated
by comparing the amplitude of the transmitted part of the different scattering states
as it reflects the transmission probability. For the Au contact it is clear that
only $\psi_{1}^{\sigma}$ penetrates significantly the tunnel gap between the apex atoms.

For the Pt junctions on the other hand the total transmission has significant contributions
from three eigenchannels in the contact regime, cf.~Fig.~\ref{FreqTrans_vs_L:fig}(d). As revealed in Fig.~\ref{Eigenchannels:fig}(b)
the symmetry of the most transmitting channel is of $\sigma$-type while the following two are of $\pi$-type
with a nodal plane through the symmetry axis.
This multichannel nature reflects the partially filled $sd$ valence shells of Pt
\cite{BrSoJa.97.Conductanceeigenchannelsin,ScAgCu.98.signatureofchemical,CuYeMa.98.MicroscopicOriginof,AgYeva.03.Quantumpropertiesof}.

\subsection{Au contacts}
\label{sec3-1}

\subsubsection{$\delta G(V)$ and $\delta \dot{S}(V)$ characteristics}
\label{sec3-1-a}

%
\begin{figure}[ht]
\centering
   \includegraphics[width=\columnwidth]{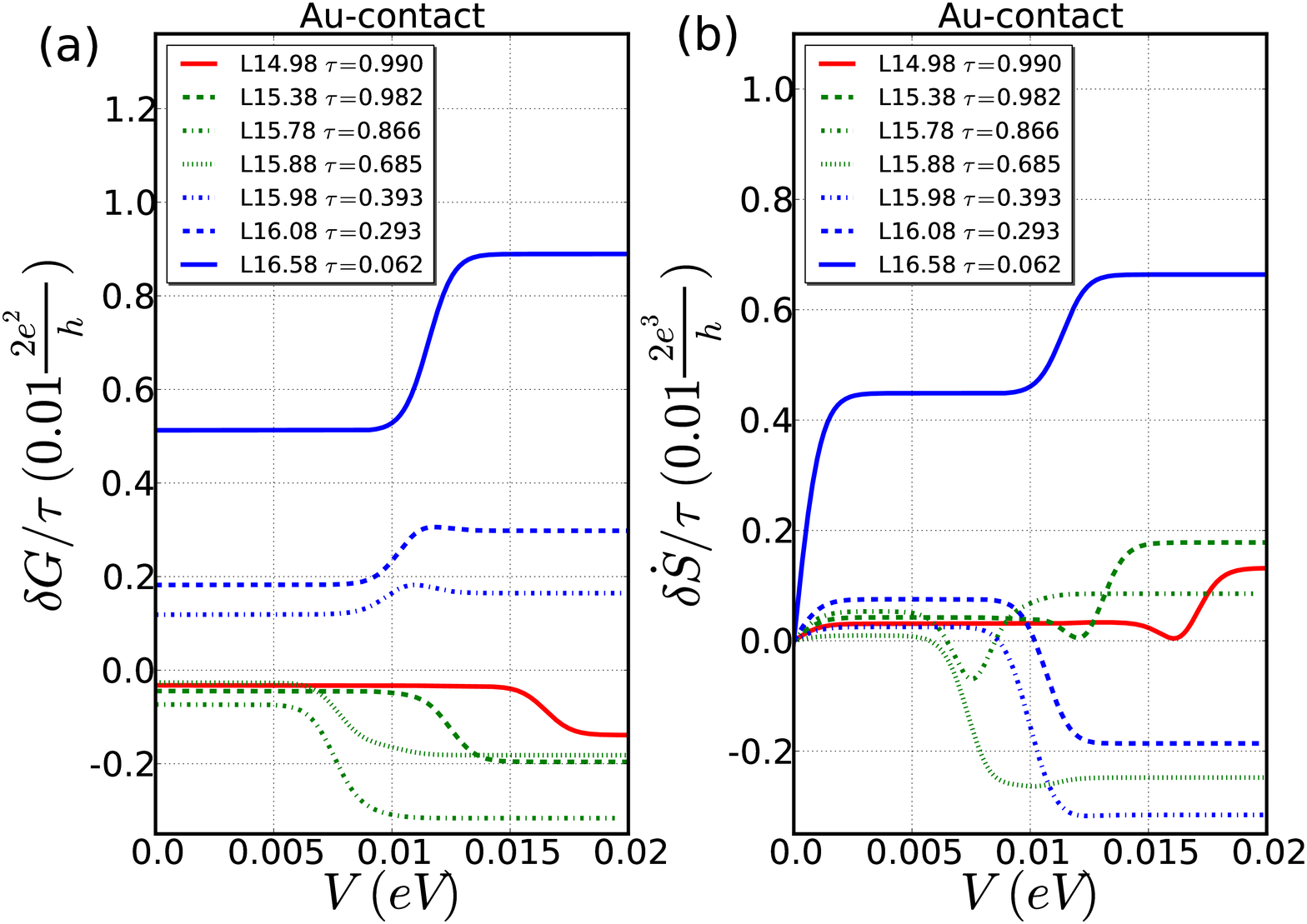}
\caption{\label{Au-contact_Characteristics:fig} 
(Color online) Inelastic conductance and noise corrections for Au atomic point contacts
with different electrode separations $L$ in the regime of equilibrated phonons.
(a) Conductance corrections $\delta G / \tau (V)$ induced by e-ph interactions as a function of voltage $V$.
(b) Derivative of the shot noise with respect to voltage $\delta \dot{S} / \tau (V)$ induced by e-ph interactions.
For each geometry six eigenmodes are considered as only the two apex atoms are vibrating,
cf.~Fig.~\ref{FreqTrans_vs_L:fig}(a).
The calculations are performed at $T=4.2\mbox{ K}$.}
\end{figure}
%
%
Using the methodology presented in Sec.~\ref{sec2-3} we proceed by studying the inelastic effects in the transport
through the considered Au atomic point contacts.
Figure~\ref{Au-contact_Characteristics:fig} shows the curves obtained for the
$\delta G(V)$ and $\delta \dot{S}(V) = \partial_{V} (\delta S(V))$ characteristics upon phonon excitation,
for several electrode distances spanning the range from tunnel to contact. 

As shown in Fig.~\ref{Au-contact_Characteristics:fig}(a), for each of the considered geometries
the correction to the reduced conductance $\delta G(V)$ exhibits 
a threshold-like character around $|eV| \sim \hbar \omega_{\leftarrow \rightarrow}
\approx 10\mbox{ meV}$ corresponding to the out-of-phase longitudinal 
vibrational mode \cite{FrLoPa.07.Fromtunnelingto}. The signals from the other five modes are so small
that they are hardly visible.
For the tunneling setups ($\tau < 1/2$) the activation of phonon
emission processes above the inelastic threshold opens a new channel for conduction, thus increasing the 
conductance compared to its elastic background value.
Contrary, in the contact regime ($\tau > 1/2$) the activation of inelastic scattering processes
reduces the conductance (backscattering), \textit{i.e.}, results in a negative jump. 

Figure~\ref{Au-contact_Characteristics:fig}(b) shows
the corresponding $\delta \dot{S}(V) = \partial_{V}(\delta S(V)$) characteristics,
which also exhibit a threshold response for voltages close to the phonon
frequency of the ``${\leftarrow \rightarrow}$'' mode. But in contrast to the conductance curves, the finite temperature effect is 
not only to smoothen the jump but also to produce some small downturn in the vicinity of the
inelastic threshold \cite{Haupt:2009,Haupt:2010}.
The sign of the jumps 
$\Delta \dot{S}_{\leftarrow \rightarrow}$ 
is consistent with the predictions based on ``single-level, single-mode'' models in 
Refs.~\cite{Schmidt:2009,Avriller:2009,Haupt:2009}, \textit{i.e.}, that it is negative only in the region $0.15 \leq \tau \leq 0.85$ and positive elsewhere. The onset of a negative inelastic correction to shot noise 
is not particularly intuitive. It was recently observed experimentally in shot noise measurements performed on
Au nano-junctions and explained in terms of correlated two-electron processes mediated by Pauli principle (Pauli blocking)
and e-ph interactions \cite{Kumar:2011}.

%
\begin{figure}[ht]
\centering
     \includegraphics[width=\columnwidth]{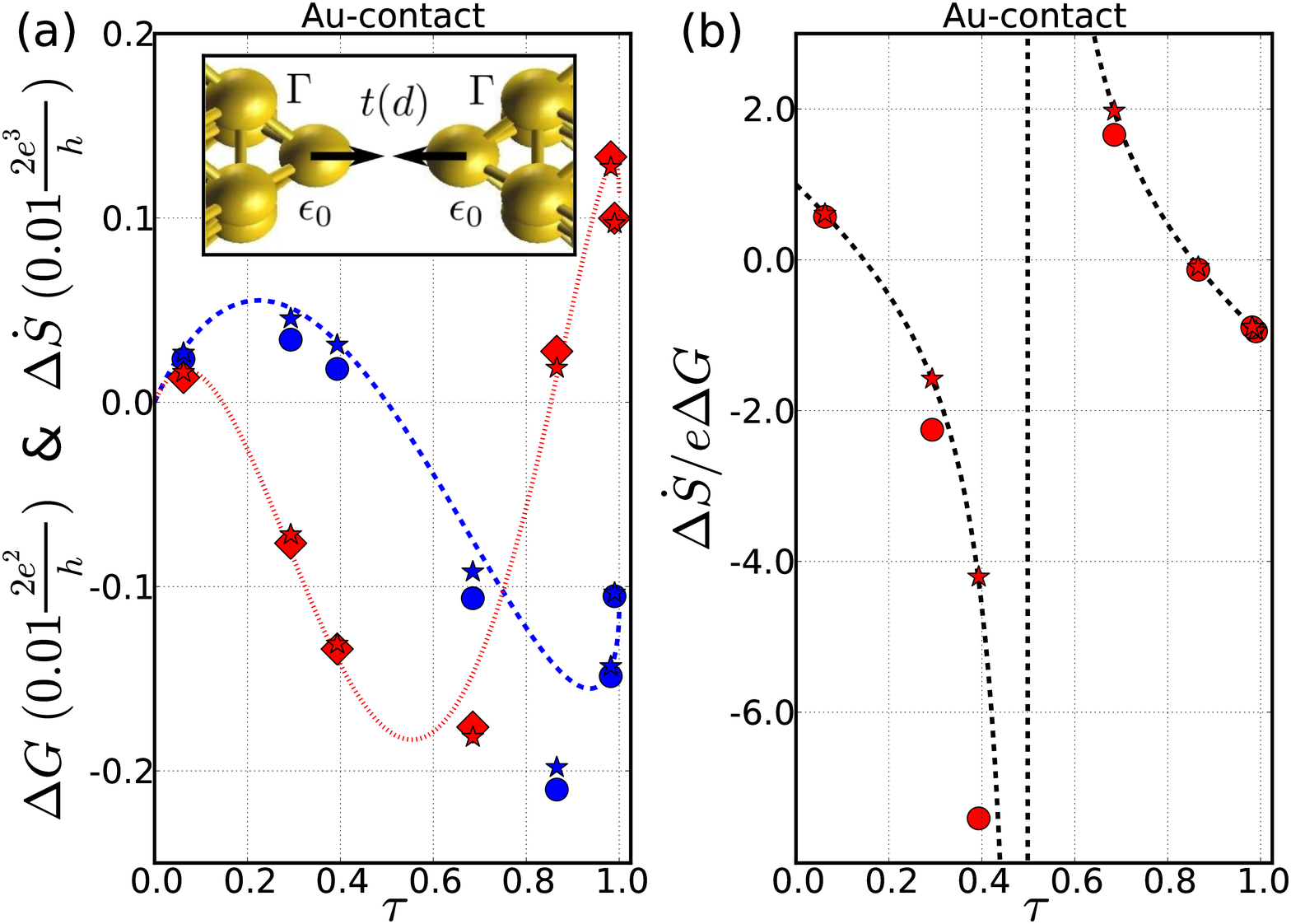}
\caption{\label{Au-contact_Jump_Analysis:fig} 
(Color online) Analysis of inelastic features for Au atomic point contacts with varying electrode separation
$d$ shown as a function of the total transmission factor $\tau$.
(a) Absolute values of the total jump in the conductance $\Delta G$ (blue circles) and 
derivative of shot noise versus voltage $\Delta \dot{S}$ (red diamonds) 
at zero temperature.
Blue-dashed and red-dotted curves are analytic results for $\Delta G$ and 
$\Delta \dot{S}$, respectively, obtained within the corresponding two-site tight-binding model of the vibrating nano-junction
(shown in inset).
(b) The ratio $\Delta \dot{S} / e\Delta G$ (red circles) as a function of the transmission 
factor at zero temperature. The black-dashed curve represent the corresponding analytic result. 
Common to both panels: Stars correspond to the inelastic signals when retaining only the contribution 
of the longitudinal out-of-phase vibrational mode $\Delta G_{\leftarrow\rightarrow}$ and $\Delta \dot{S}_{\leftarrow\rightarrow}$.
The parameters for the two-site tight-binding model are $m_{0} =0.0167\Gamma$, $t_{0}=0.875\Gamma$, and $\epsilon_{0}=E_{F}$ (see text for details).
}
\end{figure}
%

A more direct way to appreciate these trends is shown in Fig.~\ref{Au-contact_Jump_Analysis:fig}(a).
Here the total jumps in the conductance $\Delta G = \sum_{\lambda} \Delta G_{\lambda}$
(blue circles) and derivative of shot noise versus voltage $\Delta \dot{S} = \sum_{\lambda} \Delta \dot{S}_{\lambda}$ (red diamonds)
[over a voltage range that covers all possible phonon excitations] is shown
as a function of the bare transmission $\tau$ of each considered geometry.
For comparison the corresponding jumps $\Delta G_{\leftarrow \rightarrow}$ and 
$\Delta \dot{S}_{\leftarrow \rightarrow}$ associated with just the most active ``$\leftarrow \rightarrow$'' mode is shown with stars.
All the computed data is consistent with sign changes in the inelastic correction at $\tau=1/2$ for the 
conductance and at $\tau\approx \{0.15, 0.85\}$ for the shot noise.

\subsubsection{Analytic model}
\label{sec3-1-b} 

To develop an understanding for the calculated amplitudes and sign changes for $\Delta G$
and $\Delta \dot{S}$ shown in Fig.~\ref{Au-contact_Jump_Analysis:fig}(a), we developed a simple
two-site tight-binding model of the vibrating Au contact along the lines of Ref.~\cite{FrLoPa.07.Fromtunnelingto}.
This model is more appropriate to describe our physical problem than the single-level model analyzed in 
Refs.~\cite{Schmidt:2009,Avriller:2009,Haupt:2009}.

As shown in the inset to Fig.~\ref{Au-contact_Jump_Analysis:fig}(a)
we represent each of the two apex atoms in the Au contact by a single orbital
and write the noninteracting electronic Hamiltonian of this device region as
\begin{eqnarray} 
\mathbf{H}_\mathrm{two-site}^{(0)} = 
\left[ \begin{array}{cc}
\epsilon_{0} & t(d) \\
t(d) & \epsilon_{0} \end{array} \right],
\label{eqn:Hamiltonian_2sites_model}
\end{eqnarray} 
where $\epsilon_{0}$ is the onsite energy of each orbital (chosen to be equal to the Fermi energy 
$E_{F}$ for simplicity) and $t(d)$ is the hopping term that is modulated when varying the distance $d$ between the electrodes.
The hybridization of each orbital with its metallic electrode is described by 
\begin{eqnarray} 
\mathbf{\Gamma}_L = \left[ \begin{array}{cc} \Gamma & 0 \\ 0 & 0 \end{array} \right],\qquad
\mathbf{\Gamma}_R = \left[ \begin{array}{cc} 0 & 0 \\ 0 & \Gamma \end{array} \right],
\label{eqn:Gammas}
\end{eqnarray} 
where $\Gamma$ characterizes the coupling strength to each lead.

For the dependence of $t$ on distance $d$ we adopt the simple relationship
\begin{eqnarray}
t(d)=\frac{t_{0}}{1+e^{(d-d_{0})/D}},
\label{eqn:t_dependence_with_d}
\end{eqnarray}
that interpolates between the tunneling regime ($d\gg d_{0}$) for which the hopping decreases exponentially with
distance and the contact regime ($d\approx d_{0}$) for which it decreases linearly with distance.
In Eq.\refe{eqn:t_dependence_with_d}, the parameter $d_{0}$ is the typical distance where the contact 
is formed, $t_{0}$ is an energy scale that provides the prefactor for the exponential decay in the tunnel regime
(taken to be larger than $\Gamma/2$) and
$D$ describes the size of the crossover region between the two regimes.

Within the EWBL the elastic transmission for this two-site model is given by
\begin{eqnarray}
\tau(d) = \Big(\frac{\Gamma t(d)}{(\Gamma/2)^{2}+ t(d)^{2}}\Big)^{2}.
\label{eqn:transmission}
\end{eqnarray}
As it is reasonable to consider $\Gamma$ to be independent of $d$ (and to be the largest energy 
scale of the model), we restrict
the value of the hopping to be in the physically relevant branch
$0 < t(d) \leq \Gamma/2 < t_{0}$, for which one sees that the correspondence between $\tau(d)$ and $t(d)$ is one to one,
i.e., the transmission factor $\tau(t)$ is a bijective function of the hopping spanning the range 
$0<\tau(t)\leq 1$ and simply decreases with $d$ 
because of the dependence in Eq.\refe{eqn:t_dependence_with_d}.

Finally, the two atoms are coupled to the out-of-phase longitudinal vibrational
mode $(\leftarrow\rightarrow)$ (vibrating at a frequency $\omega_{\leftarrow\rightarrow}$)
which modulates the interatomic distance $d$.
The corresponding e-ph coupling matrix $\mathbf{M}^{\leftarrow\rightarrow}$ can therefore be determined 
as the derivative of the electronic Hamiltonian with respect to $d$, \textit{i.e.},
\begin{eqnarray}
\mathbf{M}^{\leftarrow\rightarrow}[t(d)] &=& 
m_0\frac{t(d)\lbrack t_{0}-t(d) \rbrack}{\Gamma/2(t_{0}-\Gamma/2)} \left[ \begin{array}{cc}
0 & 1 \\
1 & 0 \end{array} \right].
\label{eqn:eph_dependence_with_t}
\end{eqnarray}
This expression for the e-ph coupling properly captures the physical behavior with $d$, 
namely $\mathbf{M}^{\leftarrow\rightarrow}$ is proportional to $t(d)$ in the tunnel regime (exponential dependence on hopping amplitude) and is almost constant 
in the contact regime (linear dependence on hopping amplitude). 
The coupling matrix
$\mathbf{M}^{\leftarrow\rightarrow}$ depends on two parameters,
namely $m_0$ which is the value of the e-ph coupling at unit transmission (obtained for $t(d)=\Gamma/2$) and
the hopping energy scale $t_{0}$. The dependence of the coupling matrix with the distance $d$ 
is being encoded into the one to one relation $\tau[t(d)]$ and is thus not explicit in 
Eq.\refe{eqn:eph_dependence_with_t}.

In the zero temperature limit, the inelastic corrections to the mean current 
and shot noise for this two-site model can simply be expressed as
\widetext
\begin{eqnarray}
\delta I[2e/h] (V) &\approx& \gamma(\tau;m_0,t_0) \Big{\lbrace}
2(1-\tau)eV\label{eqn:Correction_Current_Onelevel} + (1-2\tau)(eV-\hbar\omega_{\leftarrow\rightarrow})\theta(|eV|-\hbar\omega_{\leftarrow\rightarrow})\Big{\rbrace}, \\
\delta S[2e^{2}/h](V)& \approx &\gamma(\tau;m_0,t_0)  \Big{\lbrace}
2(1-\tau)(1-2\tau)|eV| \label{eqn:Correction_Noise_Onelevel}  + \big{\lbrack} 1 - 8\tau(1-\tau) \big{\rbrack} (|eV|-\hbar\omega_{\leftarrow\rightarrow})
\theta(|eV|-\hbar\omega_{\leftarrow\rightarrow})\Big{\rbrace},
\end{eqnarray}
\endwidetext \noindent
where 
\begin{eqnarray}
\gamma(\tau;m_0,t_0)  = \tau\Big{(}\frac{m_{0}\lbrack t_{0}-t(\tau) \rbrack}{\Gamma/2(t_{0}-\Gamma/2)}\Big{)}^2,
\label{eqn:eph_effective_coupling} 
\end{eqnarray}
is an effective e-ph coupling constant that depends on all the parameters describing 
the electronic and vibrational structure of the junction, namely $\tau$, $m_0$ and $t_0$. 

The comparison of our \textit{ab-initio} results with Eqs.\refe{eqn:Correction_Current_Onelevel}-(\ref{eqn:Correction_Noise_Onelevel}) is shown
in Fig.~\ref{Au-contact_Jump_Analysis:fig}(a). The blue-dashed and red-dotted lines correspond to the 
results for the analytical jumps $\Delta G$ and $\Delta \dot{S}$, respectively.
We modulated the transmission factor $\tau$ by decreasing the value of the 
hopping term from $t(d)=\Gamma/2$ (in the contact regime $\tau=1$) to $t(d) \approx 0$
(in the tunnel regime $\tau \approx 0$).
We found that a reasonable (although not perfect) fit to the \textit{ab-initio} data points
could be achieved by fixing the two independent parameters
$m_0=0.0167\Gamma$ and $t_{0}=0.875\Gamma$. It is interesting to notice that the 
simple analytical model of Refs.~\cite{Schmidt:2009,Avriller:2009,Haupt:2009} fails to reproduce both the shape and amplitude 
of the curves in the full range of transmissions $\tau\in [0,1]$ (not shown here),  
mainly because the e-ph coupling strength changes when varying the distance between the electrodes in a way
qualitatively provided by Eq.\refe{eqn:eph_dependence_with_t}.
We also note that the analytic results in Fig.~\ref{Au-contact_Jump_Analysis:fig}(a) display 
clear asymmetries with respect to $\tau = 1/2$. This is because the situations $\tau \rightarrow 0$ (tunnel limit)
and $\tau \rightarrow 1.0$ (ballistic limit) are physically very different (this asymmetry is also
present in the single-level models of Refs.~\cite{Schmidt:2009,Avriller:2009,Haupt:2009}).
Moreover, the inelastic corrections, as given by Eqs.\refe{eqn:Correction_Current_Onelevel}-\refe{eqn:Correction_Noise_Onelevel}-\refe{eqn:eph_effective_coupling}, have nontrivial
dependences on $\tau$. For the chosen model parameters we find extrema in $\Delta G$ around 
$\tau\approx \{0.22,0.94\}$ and in $\Delta \dot{S}$ around $\tau\approx \{0.07,0.55,0.99\}$.

A remarkable feature of the two-site tight-binding model is
that the ratio of $\Delta \dot{S}[2e^{3}/h]$ to $e\Delta G[2e^{2}/h]$ is a universal function of $\tau$ and is independent
of the effective e-ph coupling strength $\gamma(\tau;m_0,t_0)$, \textit{i.e.},
\begin{eqnarray}
\frac{\Delta \dot{S} [2e^{3}/h]}{e\Delta G [2e^{2}/h]} &=& \frac{ 1 - 8\tau( 1 - \tau ) }{ 1 - 2\tau }  .
\label{eqn:Ratio_Jump_S_G} 
\end{eqnarray}  
This result is also found for the single-level model of Refs.~\cite{Schmidt:2009,Avriller:2009,Haupt:2009}
as a common prefactor containing the details of the electronic structure cancels out. 
Figure~\ref{Au-contact_Jump_Analysis:fig}(b) shows that our \textit{ab-initio} data follows quantitatively 
the analytic results for the ratio $\Delta \dot{S}/e \Delta G$ (black-dashed curve). The agreement with the analytical result
is even better when considering only the contribution from the out-of-phase longitudinal vibrational mode
to the inelastic signals, \textit{i.e.},  $\Delta G^{\leftarrow\rightarrow}$ and $\Delta \dot{S}^{\leftarrow\rightarrow}$
(red stars).

\subsection{Pt-contacts}
\label{sec3-2}

\subsubsection{$\delta G(V)$ and $\delta \dot{S}(V)$ characteristics}
\label{sec3-2-a}

%
\begin{figure}[ht]
\centering
   \includegraphics[width=\columnwidth]{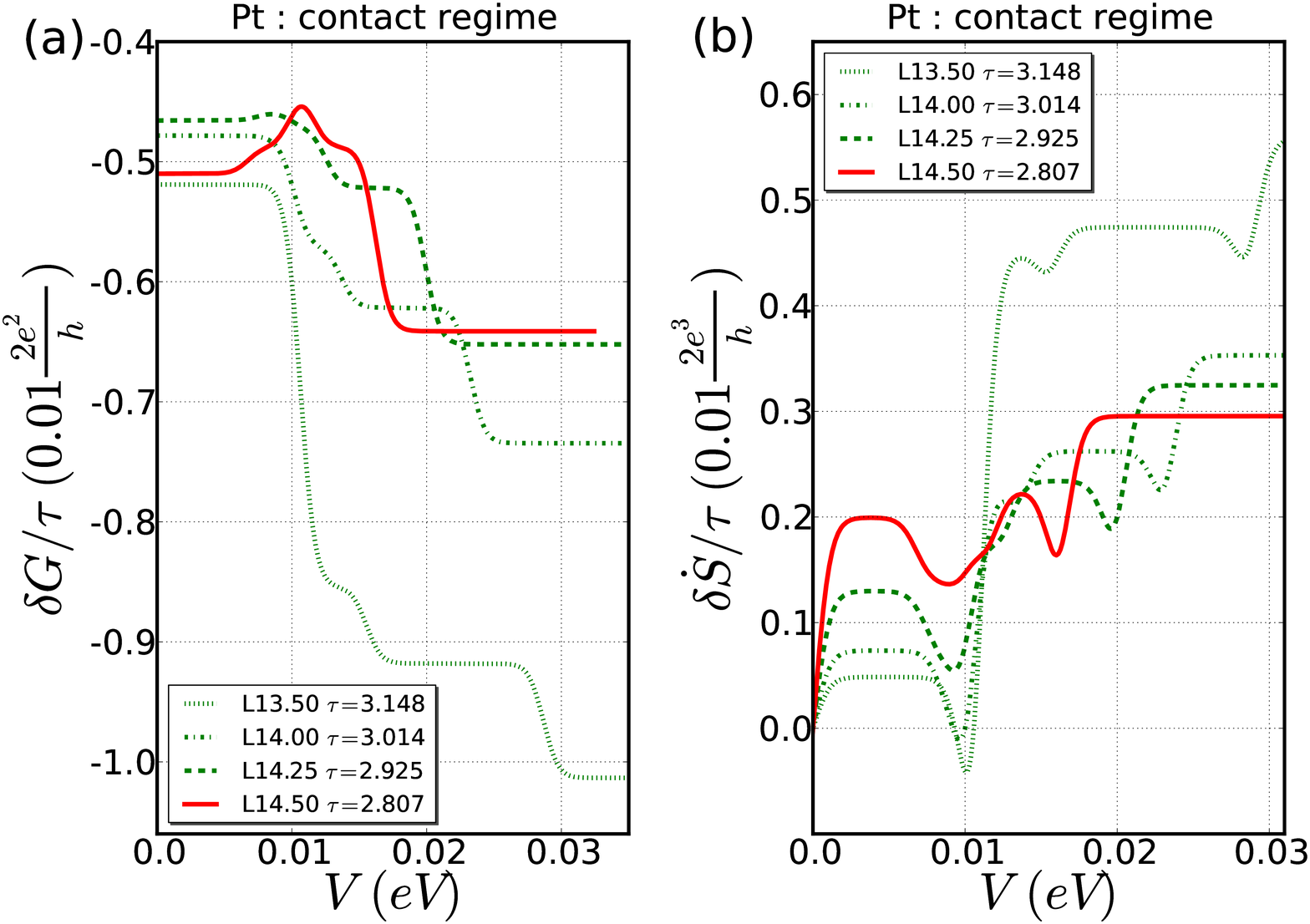}
   \includegraphics[width=\columnwidth]{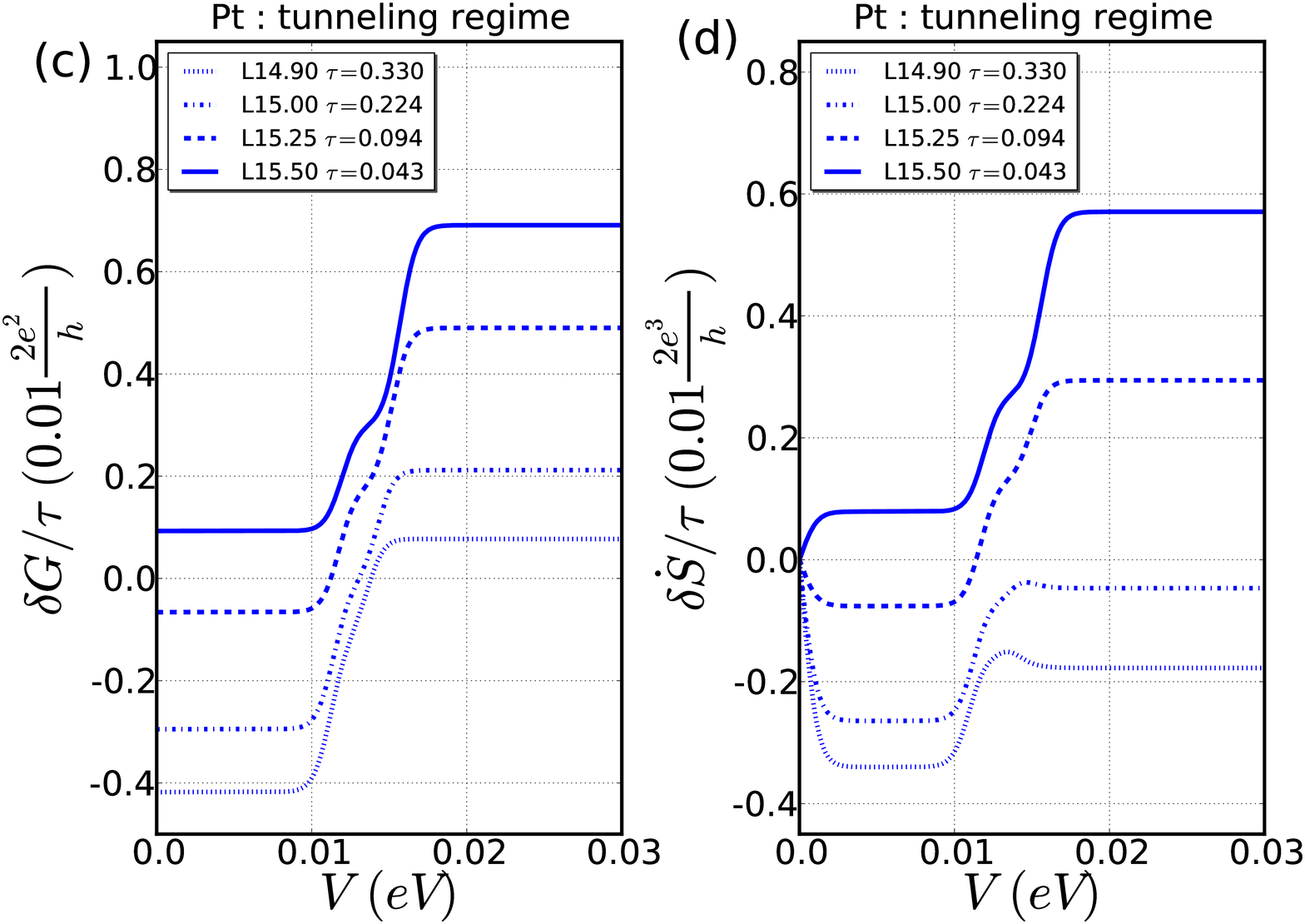}
\caption{\label{Pt-contact_Characteristics_Zoom:fig} 
(Color online) Inelastic conductance and noise corrections for Pt atomic point contacts
with different electrode separations $L$ in the regime of equilibrated phonons.
(a) Conductance corrections $\delta G / \tau (V)$ (in the contact regime) induced by e-ph interactions as a function of voltage $V$.
(b) Derivative of the shot noise with respect to voltage $\delta \dot{S} / \tau (V)$ (in the contact regime) induced by e-ph interactions.
(c)-(d) as (a)-(b) but for the geometries corresponding to the tunneling regime.
For each geometry six eigenmodes are considered as only the two apex atoms are vibrating, 
cf.~Fig.~\ref{FreqTrans_vs_L:fig}(c).
The calculations are performed at $T=4.2\mbox{ K}$.
}
\end{figure}
%
%
We present in Fig.~\ref{Pt-contact_Characteristics_Zoom:fig} the corresponding results
for Pt atomic point contacts as was given in Fig.~\ref{Au-contact_Characteristics:fig} for Au contacts.
In contrast to the Au case the electronic transport properties of Pt contacts
can no longer be understood in terms of a single conducting eigenchannel.
In fact, as seen from Fig.~\ref{FreqTrans_vs_L:fig}(d) the contact regime is 
characterized by three almost open channels, \textit{i.e.}, 
one $\sigma$-type (labelled $\psi_{1}^{\sigma}$) and two $\pi$-type (labelled $\psi_{2,3}^{\pi}$)
as visualized in Fig.~\ref{Eigenchannels:fig}(b). The fourth channel $\psi_{4}^{\sigma}$ is
included as it turns out that scattering into such closed channels are important to
understand the inelastic transport characteristics.
As for the Au contact the transmission factor decreases with $L$ and 
drops suddenly at the point where the contact (chemical bond) breaks
($\tau \approx 2.14$ for the critical geometry $L=14.80$ {\AA}). Beyond this point
the transmission drops exponentially with $d$ signalling the tunnel regime
[see Fig.~\ref{FreqTrans_vs_L:fig}(d)].

The case of the contact regime ($\tau \approx 3.0$) is shown 
in Fig.~\ref{Pt-contact_Characteristics_Zoom:fig}(a)-(b).
The inelastic features in the $\delta G(V)$ and $\delta \dot{S}(V)$ characteristics
reveal several steps associated to the excitation of transverse and 
longitudinal vibrational modes.
In this transport regime, the jumps $\Delta G^{(\lambda)}$ ($\Delta \dot{S}^{(\lambda)}$) are not necessarily
negative (positive)---\textit{i.e.}, dominated by inelastic backscattering processes---as 
expected for a single-channel system close to the ballistic limit. As shown in Fig.~\ref{Pt-contact_Characteristics_Zoom:fig}(a)-(b),
the sign of the jumps can be the other way around (see also the discussion in \Secref{sec3-2-b}).
When the transmission factor decreases, one enters into the tunnel regime (the junction breaks).
As shown in Fig.~\ref{Pt-contact_Characteristics_Zoom:fig}(c)-(d), two main inelastic
signals are seen and the jumps $\Delta G^{(\lambda)}$ and $\Delta \dot{S}^{(\lambda)}$ are always positive
in the case of low transmissions.

\subsubsection{Mode by mode analysis}
\label{sec3-2-b} 

\begin{figure}[ht]
\centering
   \includegraphics[width=\columnwidth]{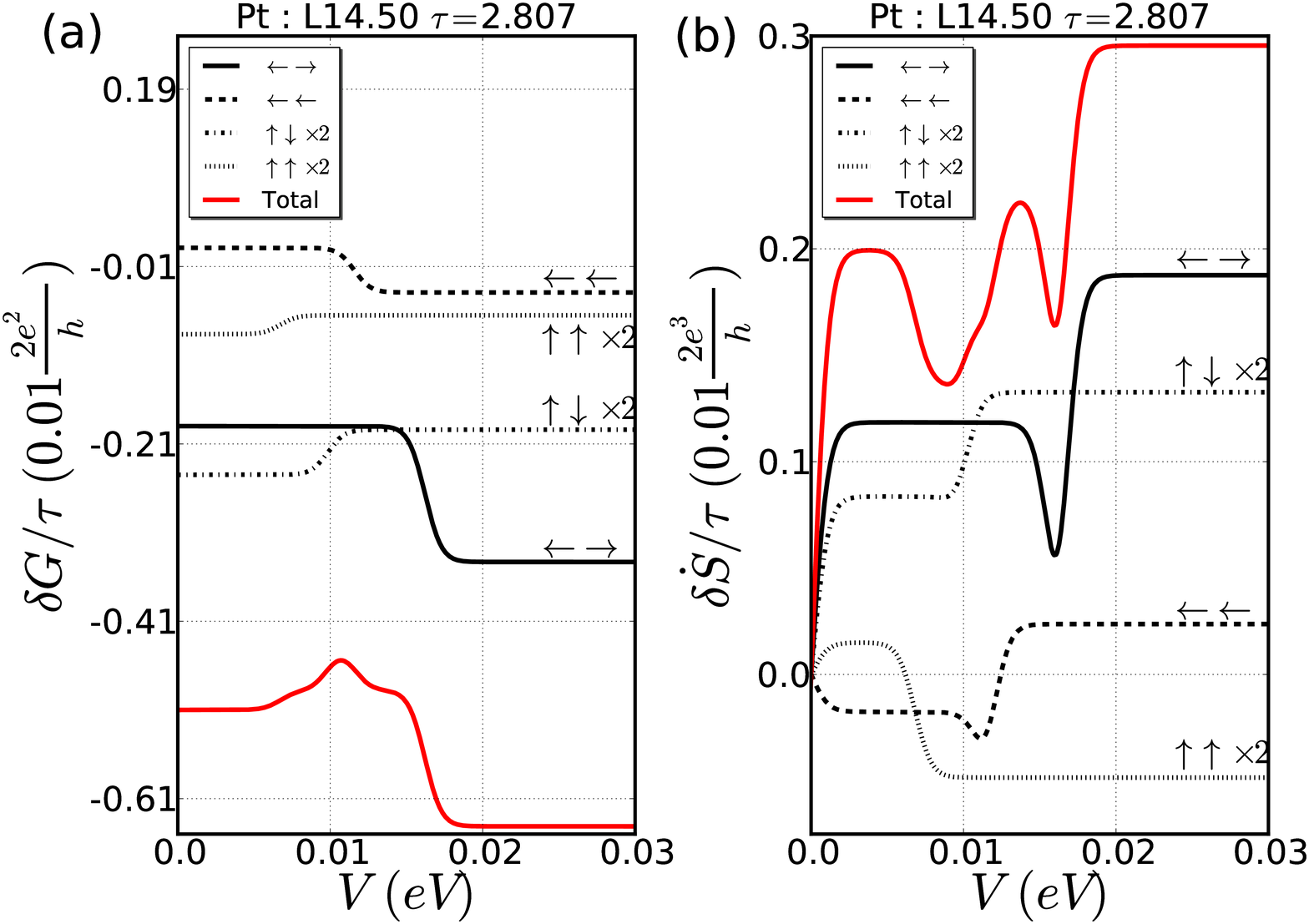}
   \includegraphics[width=\columnwidth]{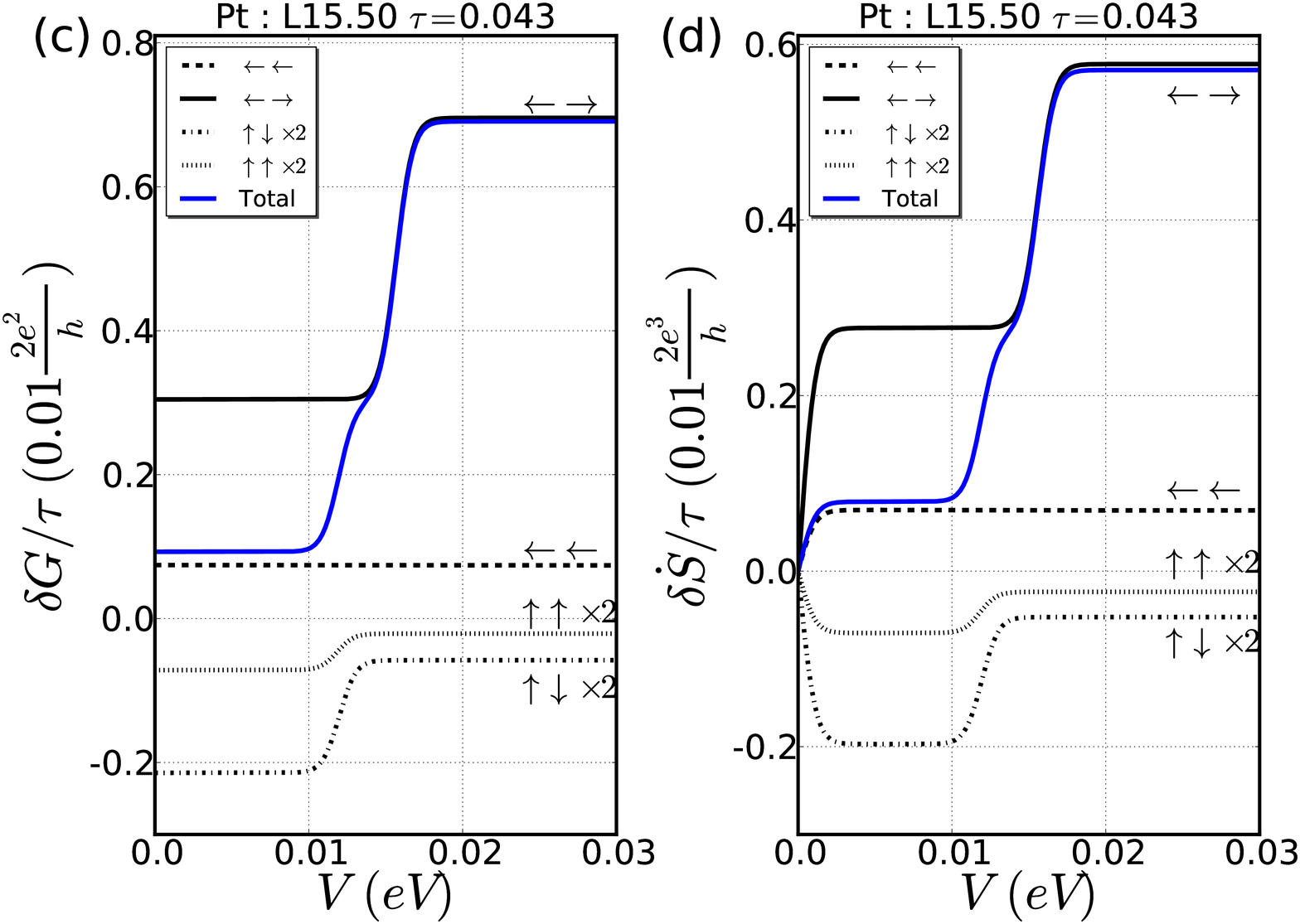}
\caption{\label{Pt-contact_Mode_by_Mode_Analysis:fig} 
(Color online) Mode by mode analysis of inelastic features for two characteristic Pt atomic point contacts in the contact
and tunneling regimes, respectively.
(a) Correction to the conductance $\delta G / \tau $  and
(b) correction to the derivative of the shot noise with respect to
voltage $\delta \dot{S} / \tau $ from the six characteristic vibrational modes 
as a function of voltage $eV$ for $L=14.50$ {\AA} (contact regime).
(c)-(d) as (a)-(b) but for $L=15.50$ {\AA} (tunnel regime).
The calculations are performed at $T=4.2\mbox{ K}$. The mode character is shown for each curve with two arrows
similar to Fig.~\ref{FreqTrans_vs_L:fig}(c).
The total characteristics $\delta G/ \tau(V)$ and $\delta \dot{S}/ \tau(V)$ (sum over all vibrational 
modes) are shown as colored plain lines.
}
\end{figure}

Due to the multichannel nature of the Pt contacts, the $\delta G(V)$ and $\delta \dot{S}(V)$ characteristics cannot be
described within the framework of the simple analytical model used for Au contacts in \Secref{sec3-1-b}.
Instead, we can gain an understanding for the characteristics by analyzing 
the contribution from each vibrational mode to the total inelastic 
signals $\delta G(V)$ and $\delta \dot{S}(V)$ for the Pt contacts. For simplicity, we restrict our analysis 
to the $L=14.50$ {\AA} geometry representative for the contact regime
[Fig.~\ref{Pt-contact_Mode_by_Mode_Analysis:fig}(a)-(b)] and to the 
$L=15.50$ {\AA} geometry representative for the tunnel regime
[Fig.~\ref{Pt-contact_Mode_by_Mode_Analysis:fig}(c)-(d)].
These two geometries are shown respectively in Fig.~\ref{Pt-contact_Characteristics_Zoom:fig}
with plain red and blue lines.

%

\begin{table}
(a) $L=14.50$ {\AA} -- contact regime\\
\begin{tabular}{l|cccc}
\hline
Mode ${\lambda}$ & ${\leftarrow\rightarrow}$ & ${\leftarrow\leftarrow}$ 
& ${\uparrow\downarrow \times 2}$ & ${\uparrow\uparrow \times 2}$ \\
\hline\hline
$\hbar\omega_{\lambda}$ [meV] & 16.2 & 11.6 & 9.8 & 6.7 \\
$\Delta G_{\lambda} [\%\Delta G]$ & 116.7 & 38.4 & -38.8 & -16.3 \\
$\Delta \dot{S}_{\lambda} [\%\Delta \dot{S}]$ & 72.0 & 43.0 & 51.1 & -66.1 \\
intra$^\mathrm{a}$: $\psi_i^L\leftrightarrow\psi_i^R$ ($i\leq 10$) [\%] & 88 & 61 & 0 & 0 \\
inter$^\mathrm{a}$: $\psi_i^L\leftrightarrow\psi_j^R$ ($i\neq j$) [\%] & 11 & 34 & 97 & 92 \\
inter$^\mathrm{b}$: $\psi^\sigma_{1,4}\leftrightarrow\psi_{2,3}^\pi$ [\%] & 0 & 0 & 70 & 66 \\
\hline
\end{tabular}
\vspace{4mm}\\
(b) $L=15.50$ {\AA} -- tunnel regime\\
\begin{tabular}{l|cccc}
\hline
Mode ${\lambda}$ & ${\leftarrow\rightarrow}$ & ${\leftarrow\leftarrow}$ 
& ${\uparrow\downarrow \times 2}$ & ${\uparrow\uparrow \times 2}$ \\
\hline\hline
$\hbar\omega_{\lambda}$ [meV] & 15.7 & 16.3 & 12.0 & 11.8 \\
$\Delta G_{\lambda} [\%\Delta G]$ & 65.4 & -0.1 & 26.2 & 8.5 \\
$\Delta \dot{S}_{\lambda} [\%\Delta \dot{S}]$ & 61.1 & -0.1 & 29.4 & 9.6 \\
intra\footnote{considering the 10 most transmitting eigenchannels}: $\psi_i^L\leftrightarrow\psi_i^R$ ($i\leq 10$) [\%] & 99 & 36 & 0 & 0 \\
inter$^\mathrm{a}$: $\psi_i^L\leftrightarrow\psi_j^R$ ($i\neq j$) [\%] & 0 & 46 & 99 & 97 \\
inter\footnote{considering only scattering among eigenchannels 1-4}: $\psi^\sigma_{1,4}\leftrightarrow\psi_{2,3}^\pi$ [\%] & 0 & 0 & 65 & 27 \\
\hline
\end{tabular}
\caption{\label{tab:1}
Mode by mode analysis of the vibrational frequencies $\hbar\omega_{\lambda}$ and jumps in the inelastic
signals ($\Delta G^{\lambda}$ and $\Delta \dot{S}^{\lambda}$) for two Pt atomic point contacts. 
The characteristic electrode separation is (a) $L=14.50$ {\AA} (contact regime) and (b) $L=15.50$ {\AA} (tunnel regime).
Relative contributions from inter- and intrachannel scattering processes to the total scattering rate are given in percent.}
%
\end{table}

Table \ref{tab:1} lists the vibrational modes and energies $\hbar\omega_{\lambda}$, the
corresponding inelastic corrections in conductance $\Delta G_{\lambda}$ and
shot noise $\Delta \dot{S}_{\lambda}$, and a decomposition of the underlying scattering
processes among the eigenchannels. 
The idea is that inelastic scattering can be understood in terms of Fermi's golden rule
where scattering occurs from occupied eigenchannel scattering states $\psi_i^L$ (for channel $i$) originating in the left 
electrode into empty scattering states $\psi_j^R$ (for channel $j$) originating in the right electrode (or vice versa,
depending on the bias polarity)
\cite{Paulsson:2008,GaSaAr.11.Simulationofinelastic}.
The total scattering rate, proportional to $\sum_{i,j}|\langle\psi_i^L|\mathbf{M}^\lambda|\psi_j^R\rangle|^2$, can 
therefore be decomposed into intrachannel ($i=j$) and interchannel
($i\neq j$) components. 

We begin our analysis with the results for the contact geometry ($L=14.50$ {\AA})
as shown in Fig.~\ref{Pt-contact_Mode_by_Mode_Analysis:fig}(a)-(b). 
The conductance $\delta G(V)$ curve
is dominated by the inelastic contribution from the longitudinal, out-of-phase vibrational 
mode (labelled $\leftarrow\rightarrow$)
with an energy quantum of $\hbar\omega_{\leftarrow\rightarrow}=16.2 \mbox{ meV}$.
As reported in Tab.~\ref{tab:1},
$\Delta G_{\leftarrow\rightarrow}$ is found to account for 116.7 \% of the overall conductance correction.
The table also reports that for this particular mode intrachannel inelastic transitions are clearly
dominant. In fact, 88 \% of the total scattering corresponds to intrachannel scattering involving 
the 10 most transmitting eigenchannels.
Furthermore, transitions between the $\sigma$-type and $\pi$-type states shown in Fig.~\ref{Eigenchannels:fig}
are symmetry forbidden (0 \% scattering) as expected for a longitudinal mode that creates
a rotationally symmetric deformation potential along the transport axis.
The e-ph coupling matrix $\mathbf{M}^{\leftarrow\rightarrow}$ is therefore essentially diagonal in the eigenchannel basis.
Finally, since scattering from this mode is essentially intrachannel involving only the three most transmitting eigenchannels, 
its effect can be rationalized in terms of a simple summation over three independent single-channel models \cite{Schmidt:2009,Avriller:2009,Haupt:2009}
where the resulting jump $\Delta G_{\leftarrow\rightarrow}$ is expected to be negative and
$\Delta \dot{S}_{\leftarrow\rightarrow}$ to be positive because the condition $\tau_i>0.85$ is satisfied for each of the first three eigenchannels, cf.~Fig.~\ref{FreqTrans_vs_L:fig}(d).
Indeed this is consistent with the numerics in Fig.~\ref{Pt-contact_Mode_by_Mode_Analysis:fig}(a)-(b).

As for the remaining vibrational modes the contributions to $\delta G(V)$ are rather 
small compared to the out-of-phase longitudinal mode $(\leftarrow\rightarrow)$ as
quantified in Tab.~\ref{tab:1} \footnote{We note that the observation of transverse modes
in the conductance signal was reported in Ref.~\cite{BoEdSc.09.Point-contactspectroscopyaluminium}
for multichannel atomic-size contacts of aluminum. As in our case
the excitation of transverse modes appeared as a positive correction to $\delta G(V)$, 
while the excitation of longitudinal modes appeared as a negative correction in the contact regime.}.
However, their contributions in the $\delta \dot{S}(V)$  
characteristics are---interestingly---much more pronounced. As shown in 
Fig.~\ref{Pt-contact_Mode_by_Mode_Analysis:fig}(b) we can identify several inelastic signals in the shot noise 
of comparable order of magnitude that were not clearly visible in the conductance.
An interesting case is the negative jump $\Delta \dot{S}_{\uparrow\uparrow}$ due to the excitation of the
in-phase transverse vibrational modes ($\uparrow\uparrow$, doubly degenerate).
Due to symmetry this mode does not allow for intrachannel scattering (0 \%)
and its effect can therefor not be rationalized in terms of single-channel models \cite{Schmidt:2009,Avriller:2009,Haupt:2009}.
In fact, this highlights that the e-ph coupling matrix $\mathbf{M}^{\uparrow\uparrow}$ is essentially off-diagonal in the eigenchannel basis.
Indeed the jump $\Delta \dot{S}_{\uparrow\uparrow}$ is negative despite that $\tau_i>0.85$ for the 
three most transmitting eigenchannels, contrary to the understanding derived from a single-channel picture.
The clear signature in the shot noise from the $\uparrow\uparrow$ mode 
is therefore a prominent demonstration that information about the e-ph coupling may be extracted
from noise measurements despite that the mode is essentially passive in the 
corresponding conductance characteristics.

%
\begin{figure*}[ht]
\centering
     \includegraphics[width=\textwidth]{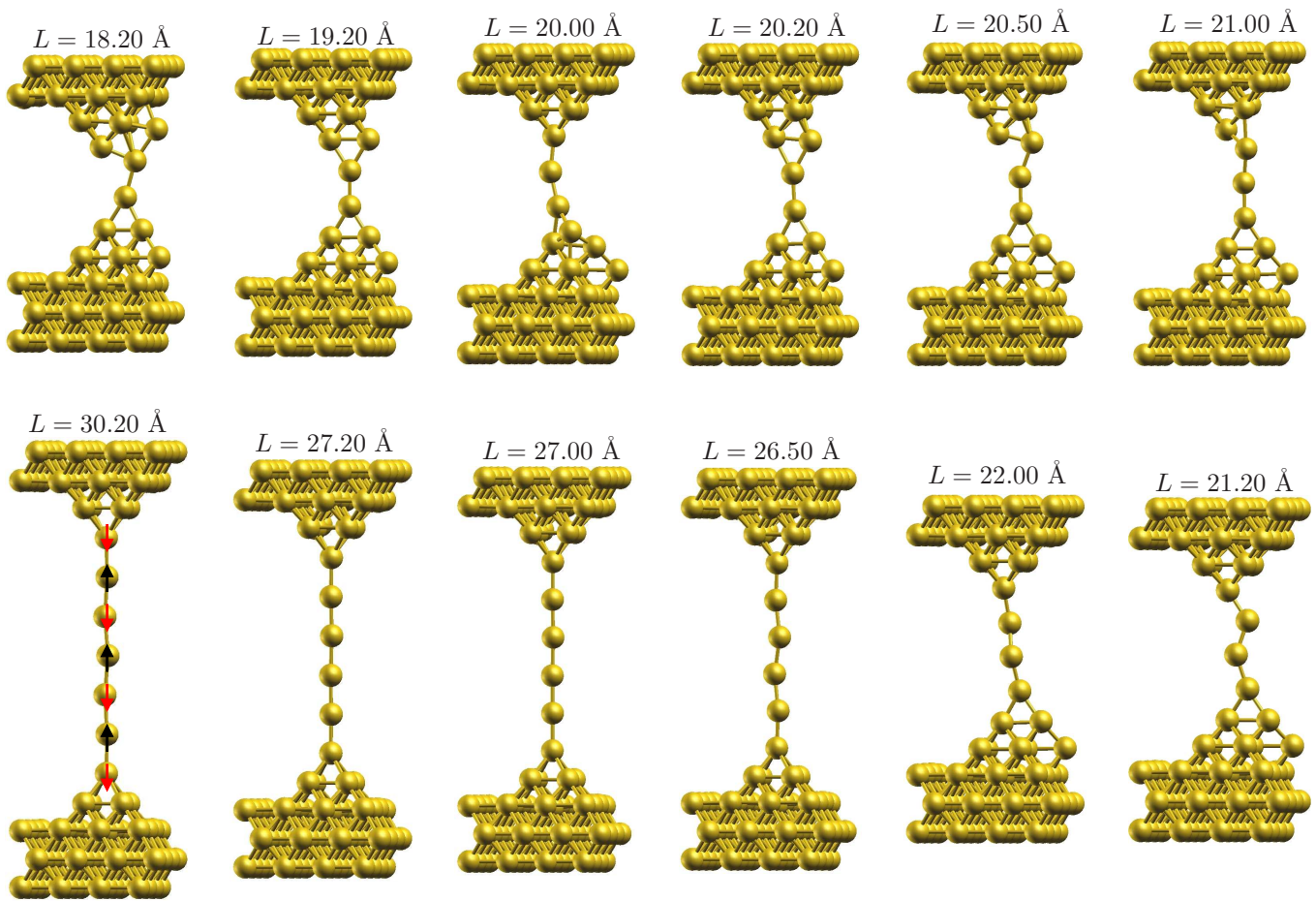}
\caption{\label{Au-nanowire:fig} 
(Color online) Stable Au atomic chain geometries considered in the first-principles transport
calculations. The characteristic electrode separation $L$, measured between the second-topmost
surface layers, is indicated over each geometry.
The active out-of-phase longitudinal (alternating bond length)
vibrational mode is sketched qualitatively with arrows in one case.}  

\end{figure*}
%

We complete our analysis by considering also a representative case in the tunneling regime
($L=15.50$ {\AA}) shown in Fig.~\ref{Pt-contact_Mode_by_Mode_Analysis:fig}(c)-(d).
The $\delta G(V)$ and $\delta \dot{S}(V)$ characteristics both exhibit positive jumps of comparable 
heights. The dominant inelastic features are located at voltages corresponding to
$\hbar\omega_{\leftarrow\rightarrow}=15.7 \mbox{ meV}$ and $\hbar\omega_{\uparrow\downarrow}=12.0 \mbox{ meV}$
that are associated with the excitation of the out-of-phase 
longitudinal mode $(\leftarrow\rightarrow)$ and of the two degenerate out-of-phase transverse modes
$(\uparrow\downarrow\times 2)$, respectively.
According to Tab.~\ref{tab:1} the $\leftarrow\rightarrow$ mode
again gives rise only to intrachannel transitions ($99\%$ of the total scattering involves intrachannel scattering for the 10 most transmitting eigenchannels). 
The e-ph coupling matrix $\mathbf{M}^{\leftarrow\rightarrow}$ is therefore almost diagonal in the eigenchannel basis
and the physics can again be understood in terms of a summation over independent channels.
The fact that the jumps $\Delta G_{\leftarrow\rightarrow}$ and $\Delta \dot{S}_{\leftarrow\rightarrow}$ are both positive
is consistent with the picture from single-level models as the condition $\tau_i<0.15$ is satisfied for all eigenchannels,
cf.~Fig.~\ref{FreqTrans_vs_L:fig}(d). 

Considering the out-of-phase transverse modes $(\uparrow\downarrow\times 2)$ 
we again find that the main inelastic processes  
are interchannel transitions (see Table ~\ref{tab:1}) and therefore that
the e-ph coupling matrix $\mathbf{M}^{\uparrow\downarrow \times 2}$  
is off-diagonal in the eigenchannel basis. 
In essence, transitions occur between the $\sigma$-type and $\pi$-type states shown in Fig.~\ref{Eigenchannels:fig}
as expected for a transversal mode due to symmetry.
Again, single-level models
are inadequate to describe the physics and therefore do not predict the sign of
the jumps $\Delta G_{\uparrow\downarrow}$ and $\Delta \dot{S}_{\uparrow\downarrow}$. 
According to our numerics we find that both jumps are positive.  

\section{Results for Au atomic chains}
\label{sec4}

Au atomic chains constitute benchmark systems for the exploration of inelastic signatures 
in the transport characteristics
\cite{AgUnRu.02.Onsetofenergy,FrBrLo.04.InelasticScatteringand,ViCuPa.05.Electron-vibrationinteractionin,VeMaAg.06.Universalfeaturesof,Frederiksen:2007}
and most recently the effect of phonon emission in the electronic shot noise was 
reported for the first time \cite{Kumar:2011}.
To complete our study we present in this section realistic calculations of inelastic signals 
in the $\delta G(V)$ and $\delta \dot{S}(V)$ characteristics for a set of Au atomic chain geometries,
shown in Fig.~\ref{Au-nanowire:fig}, corresponding to different lengths and strain conditions.

%
%
\begin{figure}[ht]
\centering
   \includegraphics[width=\columnwidth]{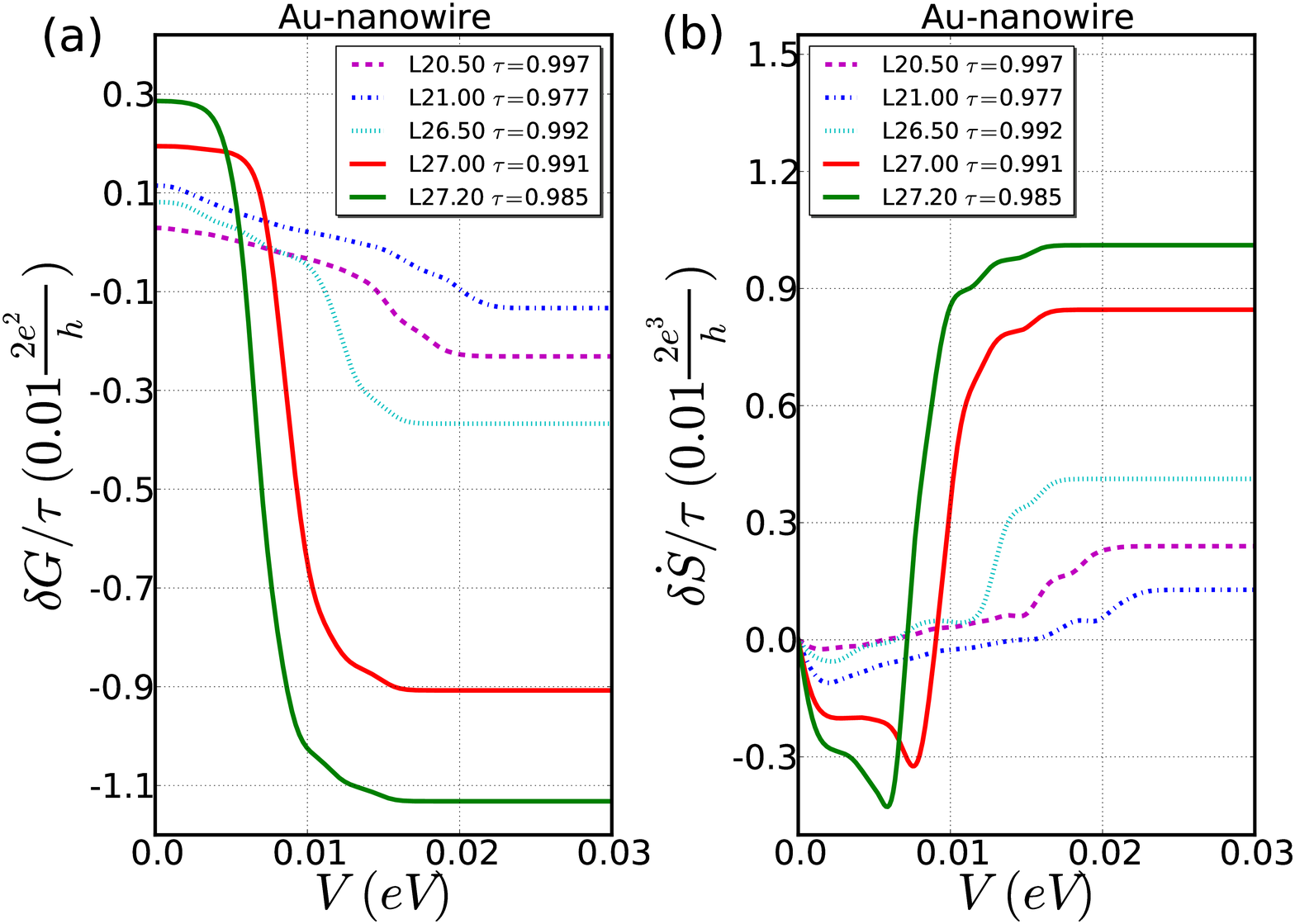}
   \includegraphics[width=\columnwidth]{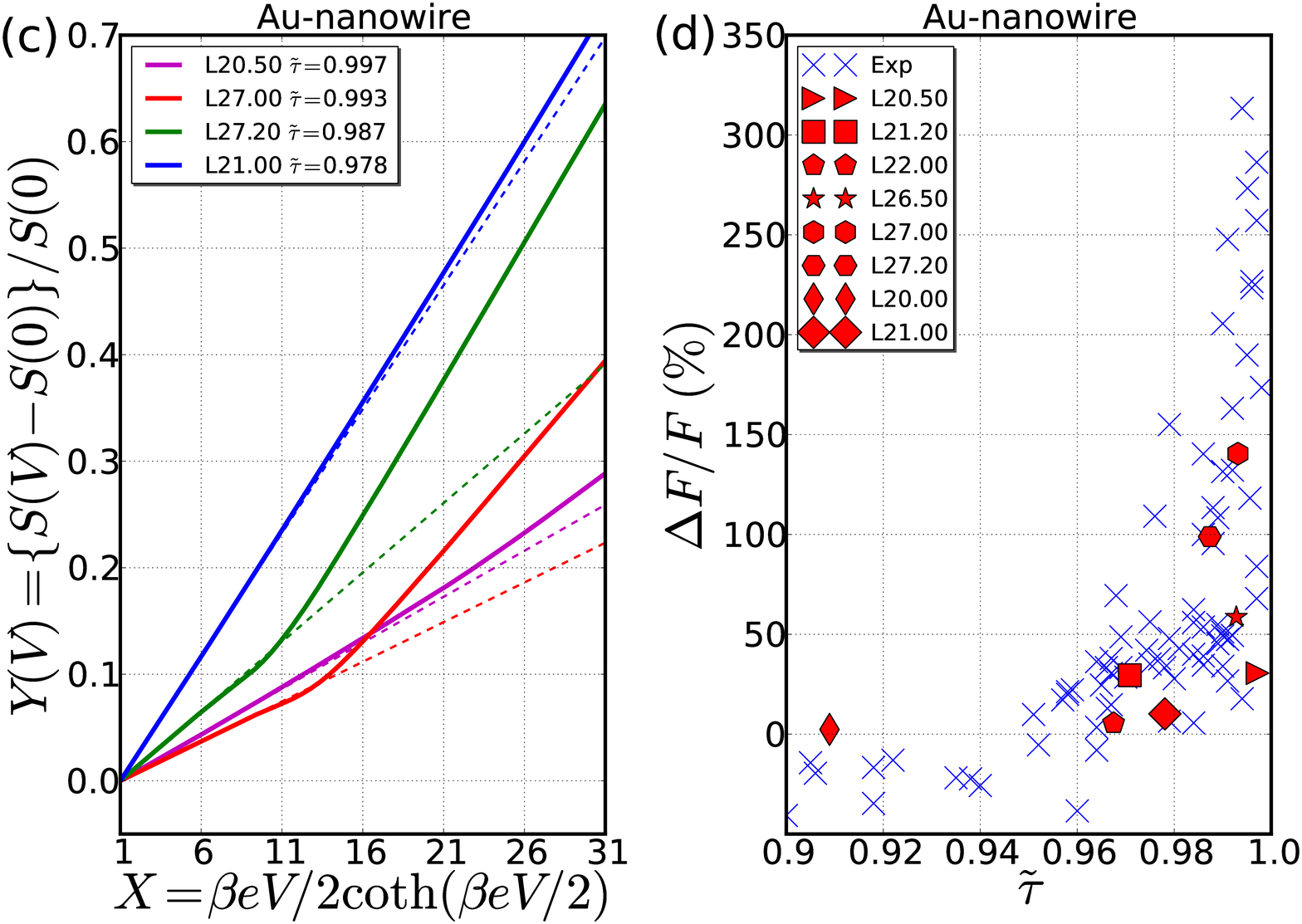}
\caption{\label{Au-nanowire_Characteristics:fig} 
(Color online) Inelastic conductance and noise characteristics for Au atomic chains
shown in Fig.~\ref{Au-nanowire:fig} in the regime of equilibrated phonons.
(a) $\delta G / \tau (V)$ and (b) $\delta \dot{S} / \tau (V)$ 
characteristics computed at $T=4.2\mbox{ K}$. 
(c) Total shot noise $S(V) = S_{0}(V) + \delta S(V)$ expressed in the reduced scale 
$Y(V) = ( S(V) - S(V=0) ) / S(V=0)$ as a function of the reduced parameter 
$X(V) = \beta eV/2 \coth( \beta eV/2 )$ computed at $T=4.2\mbox{ K}$.
The dashed curves are the extrapolation of the low voltage reduced noise characteristics
$Y_{-}(X) = F_{-}(X-1)$, with $F_{-}$ the Fano factor (see the text for details).
(d) The relative jump of Fano factor $\Delta F/F = ( F_{+} - F_{-} ) / F_{-}$ (red points) as a function of the
renormalized transmission factor 
$\tilde{\tau} \approx \tau + \sum_\lambda\mbox{Tr}\lbrace \mathbf{T}^\mathrm{(0)}_\lambda\rbrace$, as extracted
numerically from the $Y(X)$ characteristics. For comparison, the blue crosses are the experimental results
of Kumar \textit{et al.} ~\cite{Kumar:2011}.
}
\end{figure}

Our calculations were carried out in the same way as described in Sec.~\ref{sec3} for
the Au and Pt atomic contacts, except that now we allow all atoms bridging 
the Au(111) surfaces to vibrate,
\textit{i.e.}, the dynamical region consists of 15 vibrating atoms.
Again, for each electrode separation $L$ we relax the dynamical atoms and the 
topmost Au(111) layer until the residual forces are smaller than 0.02 eV/{\AA}.
All the structures considered here are stable in the sense that $\hbar\omega_\lambda>0$
for all vibrational modes (no imaginary frequency modes exist
in the calculations).

In order for us to explore the location of a possible sign change in $\delta \dot{S}(V)$
we intended to generate very different geometries with transmission factors spanning
as large an interval as possible. For all the considered structures, shown
in Fig.~\ref{Au-nanowire:fig}, we obtained the range $\tau = 0.911-1.02$
which is significantly more narrow than reported in the experiments of Ref.~\cite{Kumar:2011}.
Within our treatment it appears difficult to construct geometries 
where the transmission deviates significantly from unity (unless the 
chain is broken).

The generic behavior for the Au atomic chains---such as the plain red curve for the geometry $L=27.00$ {\AA} 
in Fig.~\ref{Au-nanowire_Characteristics:fig}(a) (for which $\tau = 0.991$)---is a main
inelastic threshold in the conductance curve $\delta G(V)$ located at voltages $|eV| \approx 10 \mbox{ meV}$.
The corresponding negative jump in conductance is due to inelastic backscattering events associated 
with the activation of the out-of-phase longitudinal (alternating bond length) vibrational
mode of the atomic chain (the zone-boundary phonon mode with wave length $q\approx 2k_{F}$)
\cite{AgUnRu.02.Onsetofenergy,FrBrLo.04.InelasticScatteringand}. This active mode is
sketched qualitatively with arrows in Fig.~\ref{Au-nanowire:fig}.
The size of the jump is of order $1\%$, which is significantly larger 
than the corresponding signals in the case of Au contacts, cf.~Fig.~\ref{Au-contact_Characteristics:fig}(a).
This is consistent with tight-binding models \cite{FrBrLo.04.ModelingofInelastic,VeMaAg.06.Universalfeaturesof}
which have shown that the inelastic conductance correction is roughly proportional to the 
number of vibrating atoms in the chain.
When considering the noise properties one observes a positive jump of similar magnitude 
in the $\delta \dot{S}(V)$ characteristics as seen in Fig.~\ref{Au-nanowire_Characteristics:fig}(b).

In order to compare the output of our numerical calculations to the experimental results of Kumar 
\textit{et al.} ~\cite{Kumar:2011}, we represented in Fig.~\ref{Au-nanowire_Characteristics:fig}(c) the 
total shot noise characteristics $S(V) = S_{0}(V) + \delta S(V)$ expressed in the reduced scale 
$Y(V) = ( S(V) - S(V=0) ) / S(V=0)$, as a function of the reduced parameter 
$X(V) = \beta eV/2 \coth( \beta eV/2 )$. As in Ref.~\cite{Kumar:2011}, the $Y(X)$ curves are well approximated by 
piecewise linear functions.
When $eV$ is below the main inelastic threshold, the slope of the $Y(X)$ curve gives the Fano factor $F_{-}$.
Upon phonon excitation, the slope changes to $F_{+}$ thus defining a modified Fano factor (due to 
the inelastic correction to shot noise $\delta S_{inel}(V)$ [Eq.\refe{eqn:Noise_Interaction_Inelastic_Part}])
and a relative jump of 
Fano factor $\Delta F/F = ( F_{+} - F_{-} ) / F_{-}$ (see Ref.~\cite{Kumar:2011}). We represent on
Fig.~\ref{Au-nanowire_Characteristics:fig}(d) the dependence of the relative jump of 
Fano factor $\Delta F/F$ (red points) as a function of the renormalized transmission factor
$\tilde{\tau} \approx \tau + \sum_\lambda\mbox{Tr}\lbrace \mathbf{T}^\mathrm{(0)}_\lambda\rbrace$ (which coincides 
with the zero bias conductance). Each point is associated to a given geometry shown in Fig.~\ref{Au-nanowire:fig} 
and is computed by numerical differentiation of the corresponding $Y(X)$ curve. The experimental points from Kumar 
\textit{et al.} ~\cite{Kumar:2011} are shown in the range $\tau = 0.9-1.0$ as blue crosses.

We first remark that for transmission factors $\tau > 0.95$ the computed points agree quantitatively with the 
experimental ones. This exemplify the predictive power of \textit{ab initio} methods for generating 
quantitative results for transport calculations without any adjustable parameter. Although our results are 
fully consistent with the experimental data from Ref.~\cite{Kumar:2011} in the regime $\tau > 0.95$, 
we are unable to identify a sign change in the correction to the shot noise, which was reported experimentally
to occur around $\tau \approx 0.95$ instead of the expected theoretical value $\tau \approx 0.85$
(see Fig.~\ref{Au-nanowire_Characteristics:fig}(d) blue crosses).

\begin{figure}[ht]
\centering
   \includegraphics[width=\columnwidth]{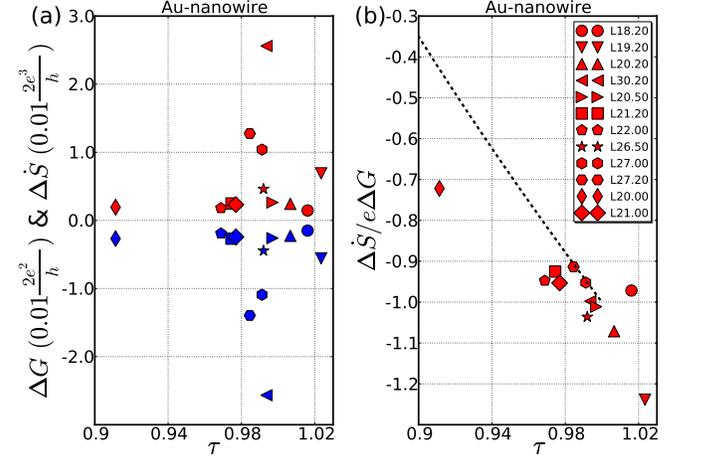}
\caption{\label{Au-nanowire_Characteristics_2:fig} 
(Color online) Inelastic conductance and noise corrections for Au atomic chains
shown in Fig.~\ref{Au-nanowire:fig} in the regime of equilibrated phonons.
(a) Total jumps $\Delta G$ (blue points) and $\Delta \dot{S}$ (red points)
at zero temperature as a function of the transmission factor.
(b) The ratio $\Delta \dot{S} / e\Delta G$ (red points) as a function of the transmission 
factor. The  black dashed curve is the prediction of analytical models for the ratio, cf.~Eq.\refe{eqn:Ratio_Jump_S_G}.
}
\end{figure}

In order to try to extrapolate the position of the crossover as obtained in the output 
of our calculations, we summarize in Fig.~\ref{Au-nanowire_Characteristics_2:fig}(a)
the dependence of the size of the total jumps $\Delta G$ and $\Delta \dot{S}$ as a function of the transmission
factor $\tau$ of the Au atomic chains. This plot clearly shows
that statistically the jumps in conductance $\Delta G$ (blue points) and 
shot noise $\Delta \dot{S}$ (red points) have opposite sign
and very similar magnitude.
The more detailed correlation between $\Delta G$ and $\Delta \dot{S}$ can be characterized by 
considering the
ratio $\Delta \dot{S}/e\Delta G$ as shown in Fig.~\ref{Au-nanowire_Characteristics_2:fig}(b). 
Here it is observed that the absolute value of the ratio tends to increase with $\tau$. However,
the computed ratios (red points) do not fall exactly on the prediction 
of analytical models (black dashed curve) such as the result in Eq.\refe{eqn:Ratio_Jump_S_G}.
Rather it appears as if the analytics is an upper bound to the computed ratios \footnote{
We also studied a tight-binding model of an $N$-site atomic chain that couples to the longitudinal
vibrational mode with alternating bond-length character. 
For an even number of sites ($N=2p$) the model has the same output as the simple Eq.\refe{eqn:Ratio_Jump_S_G}, whereas 
for odd number of sites ($N=2p+1$) we observed a different qualitative behavior of the inelastic signals where
the total jump $\Delta G$ is always negative (independent of the transmission $\tau$) and
$\Delta \dot{S}$ changes sign at $\tau=0.5$ instead of at $\tau\approx\{0.15,0.85\}$.
We assign this even-odd effect with $N$
to the vanishing vertex correction to the shot-noise due to the symmetry of the particular 
electron-phonon matrix elements.}.
The data that agrees the best with the analytic ratio corresponds to the cases of 
even-numbered, linear atomic chains $L=27.00$ {\AA} and $L=27.20$ {\AA}.
%
%

Our data in Fig.~\ref{Au-nanowire_Characteristics_2:fig}(b) actually suggests
that the crossover could occur at a lower value than the $\tau \approx 0.85$ value from analytic models.
An interesting future extension of our results would be to try to include disorder-induced
conductance fluctuations in the treatment, as this effect played an important role in
the interpretation of the experiments in Ref.~\cite{Kumar:2011}. 
Such a development with a semi-empirical model of disorder could be achieved through a 
multi-scale methodology \cite{RA.2007.MultiscaleApproaches}, but at the cost of giving up
the parameter-free scheme presented here 
for simulating the inelastic transport characteristics of nanodevices.

\section{Conclusions}
\label{sec5}

We have implemented into the \textsc{Inelastica} \cite{PaFrBr.05.Modelinginelasticphonon,Frederiksen:2007,Inelastica} code
a methodology \cite{Komnik:2006,Haupt:2010} that allows to compute quantitatively inelastic shot noise signals from first principles.
The method was illustrated for Au and Pt atomic point contacts as well as Au atomic chains 
described at the atomistic level.
We showed that the Au contact constitutes a benchmark system that can be
understood in terms of the ``single level, single vibrational mode'' models \cite{Schmidt:2009,Avriller:2009,Haupt:2009}.
To be quantitative we rationalized computed inelastic corrections to the conductance and shot noise characteristics
in terms of a ``two-site'' tight-binding model for the vibrating Au contacts.
However, in the case of Pt contacts the multichannel nature complicates the physics and
simple analytic models were shown to be inadequate. Interestingly, we found that
for Pt contact the coupling to transverse vibrational modes gives rise to 
features in the inelastic shot noise signals (originating in interchannel scattering processes)
that are not readily visible in the conductance. 
This opens up an alternative approach to characterize the e-ph couplings and the underlying multichannel 
electronic transport problem.

We also analyzed Au atomic chain configurations in order to compare directly with the experimental 
results of Ref.~\cite{Kumar:2011} for the inelastic effects in the shot noise. While the simulated
shot noise behavior was found to agree quantitatively with the experiment in the ballistic regime 
($0.95 < \tau \approx 1$) we were unable to identify the sign change in the shot noise correction 
that was reported in Ref.~\cite{Kumar:2011}.

The techniques for characterizing inelastic effects in shot noise appears as a promising way to
gain deeper insight into transport properties of nanoscale devices. It is our hope that the
methodology presented in this paper will stimulate further research in this direction.

\section*{Acknowledgements}
\label{sec6}
The authors are grateful to  Federica Haupt, Jan M.~van Ruitenbeek, and Alfredo Levy Yeyati for their
careful reading and comments on the manuscript.

\bibliography{biblioIETS}

\end{document}